\documentclass[trackchanges, default]{aastex7}

\newcommand\um{$\mu$m}
\usepackage{graphicx}
\usepackage{longtable}
\usepackage{subcaption}

\begin{document}

\title{Cosmic Noon Galaxies in the Hubble Ultra-Deep Field using MIRI Wide-Field Slitless Spectroscopy}

\author[orcid=0000-0002-7612-0469,sname='Sarah Kendrew']{Sarah Kendrew}
\affiliation{European Space Agency, ESA Office, Space Telescope Science Institute, 3700 San Martin Drive, Baltimore MD 21231, USA}
\email[show]{skendrew@stsci.edu}  
\author[orcid=0000-0002-3005-1349]{G\"{o}ran \"{O}stlin}
\affiliation{The Oskar Klein Centre, Department of Astronomy, Stockholm University, AlbaNova, SE-10691 Stockholm, Sweden}
\email{ostlin@astro.su.se}
\author[orcid=0000-0002-9090-4227]{Luis Colina}
\affiliation{Centro de Astrobiolog\'{i}a (CAB), CSIC-INTA, Ctra. de Ajalvir km 4, Torrej\'{o}n de Ardoz, E-28850, Madrid, Spain}
\email{colina@cab.inta-csic.es}
\author[orcid=0000-0002-5104-8245]{Pierluigi Rinaldi}
\affiliation{Department of Astronomy, The University of Texas at Austin, Austin, TX 78712, USA}
\affiliation{Cosmic Frontier Center, The University of Texas at Austin, Austin, TX 78712, USA}
\affiliation{Space Telescope Science Institute, 3700 San Martin Drive, Baltimore MD 21218, USA}
\email{pierluigi.rinaldi@austin.utexas.edu}
\author[orcid=0000-0002-7093-1877]{Javier \'{A}lvarez-M\'{a}rquez}
\affiliation{Centro de Astrobiolog\'{i}a (CAB), CSIC-INTA, Ctra. de Ajalvir km 4, Torrej\'{o}n de Ardoz, E-28850, Madrid, Spain}
\email{jalvarez@cab.inta-csic.es}
\author[orcid=0000-0002-3952-8588]{Leindert A. Boogaard}
\affiliation{Leiden Observatory, Leiden University, PO Box 9513, NL-2300 RA Leiden, The Netherlands}
\email{boogaard@strw.leidenuniv.nl}

\author[orcid=0000-0003-0470-8754]{Jens Melinder}
\affiliation{The Oskar Klein Centre, Department of Astronomy, Stockholm University, AlbaNova, SE-10691 Stockholm, Sweden}
\email{jens@astro.su.se}
\author[orcid=0000-0002-0932-4330]{John P. Pye}
\affiliation{School of Physics and Astronomy, Space Research Centre, Space Park Leicester, University of
Leicester, 92 Corporation Road, Leicester, LE4 5SP, UK}
\email{pye@leicester.ac.uk}

\author[orcid=0000-0001-6794-2519]{Almudena Alonso Herrero}
\affiliation{Centro de Astrobiolog\'{\i}a (CAB), CSIC-INTA, Camino Bajo del Castillo s/n, E-28692 Villanueva de la Ca\~nada, Madrid, Spain}
\email{aalonso@cab.inta-csic.es}
\author[orcid=0000-0001-8183-1460]{Karina I. Caputi}
\affiliation{Kapteyn Astronomical Institute, University of Groningen, 9700 AV Groningen, The Netherlands}
\email{karina@astro.rug.nl}
\author[orcid=0000-0003-2119-277X]{Alejandro Crespo G\'{o}mez}
\affiliation{Space Telescope Science Institute, 3700 San Martin Drive, Baltimore MD 21218, USA}
\email{acrespo@stsci.edu}
\author[orcid=0000-0003-4801-0489]{Macarena Garc\'{i}a Mar\'{i}n}
\affiliation{European Space Agency, ESA Office, Space Telescope Science Institute, 3700 San Martin Drive, Baltimore MD 21231, USA}
\email{maca@stsci.edu}
\author[orcid=0000-0001-9885-4589]{Steven Gillman}
\affiliation{Cosmic Dawn Center (DAWN), Denmark}
\affiliation{DTU-Space, Elektrovej, Building 328, 2800 Kgs. Lyngby, Denmark}
\email{srigi@space.dtu.dk}
\author[orcid=0000-0003-0129-2079]{Santosh Harish}
\affiliation{Space Telescope Science Institute, 3700 San Martin Drive, Baltimore MD 21218, USA}
\email{sharish@stsci.edu}
\author[orcid=0000-0001-8386-3546]{Edoardo Iani}
\affiliation{Institute of Science and Technology Austria (ISTA), Am Campus 1, 3400 Klosterneuburg, Austria}
\email{edoardo.iani@ist.ac.at}
\author[orcid=0000-0001-5710-8395]{Danial Langeroodi}
\affiliation{Kavli Institute for Cosmology, University of Cambridge, Madingley Road, Cambridge CB3 0HA, UK}
\affiliation{Cavendish Laboratory, University of Cambridge, 19 JJ Thomson Avenue, Cambridge CB3 0HE, UK}
\affiliation{DARK, Niels Bohr Institute, University of Copenhagen, Jagtvej 155A, 2200 Copenhagen, Denmark}
\email{danial.langeroodi@nbi.ku.dk}
\author[orcid=0000-0002-0690-8824]{Alvaro Labiano}
\affiliation{Telespazio UK SL for ESA, ESAC, Camino Bajo del Castillo s/n, E-28692, Villanueva de la Ca\~{n}ada, Madrid, Spain}
\email{Alvaro.LabianoOrtega@ext.esa.int}

\author[orcid=0000-0003-4528-5639]{Pablo G. P\'{e}rez-Gonz\'{a}lez}
\affiliation{Centro de Astrobiolog\'{i}a (CAB), CSIC-INTA, Ctra. de Ajalvir km 4, Torrej\'{o}n de Ardoz, E-28850, Madrid, Spain}
\email{pgperez@cab.inta-csic.es}
\author[orcid=0000-0001-5434-5942]{Paul van der Werf}
\affiliation{Leiden Observatory, Leiden University, PO Box 9513, NL-2300 RA Leiden, The Netherlands}
\email{pvdwerf@strw.leidenuniv.nl}
\author[orcid=0000-0003-4793-7880]{Fabian Walter}
\affiliation{Max-Planck-Institut für Astronomie, K\"{o}nigstuhl 17, 69117 Heidelberg, Germany}
\email{walter@mpia.de}
\author[orcid=0000-0001-7416-7936]{Gillian Wright}
\affiliation{UK Astronomy Technology Centre, Royal Observatory Edinburgh, Blackford Hill, Edinburgh EH9 3HJ, UK}
\email{gillian.wright@stfc.ac.uk}

\author[orcid=0000-0002-2554-1837]{Thomas Greve}
\affiliation{Cosmic Dawn Center (DAWN), Denmark}
\affiliation{DTU-Space, Elektrovej, Building 328, 2800 Kgs. Lyngby, Denmark}
\email{tgreve@space.dtu.dk}
\author[orcid=0000-0001-9818-0588]{Manuel G\"{u}del}
\affiliation{University of Vienna, Department of Astrophysics, T\"{u}rkenschanzstrasse 17, 1180 Vienna, Austria}
\affiliation{Institute of Particle Physics and Astrophysics, ETH Z\"{u}rich, Wolfgang-Pauli-Str 27, 8093 Z\"{u}rich, Switzerland}
\email{manuel.guedel@univie.ac.at}
\author[orcid=0000-0002-1493-300X]{Thomas Henning}
\affiliation{Max-Planck-Institut für Astronomie, K\"{o}nigstuhl 17, 69117 Heidelberg, Germany}
\email{henning@mpia.de}
\author{Pierre-Olivier Lagage}
\affiliation{AIM, CEA, CNRS, Universit\'{e} Paris-Saclay, Universit\'{e} Paris Diderot, Sorbonne Paris Cit\'{e}, 91191 Gif-sur-Yvette, France}
\email{pierre-olivier.lagage@cea.fr}
\author[orcid=0000-0002-2110-1068]{Tom P. Ray}
\affiliation{Dublin Institute for Advanced Studies, Astronomy \& Astrophysics Section, 31 Fitzwilliam Place, Dublin 2, Ireland}
\email{tr@cp.dias.ie}
\author[orcid=0000-0002-1368-3109]{Bart Vandenbussche}
\affiliation{Institute of Astronomy, KU Leuven, Celestijnenlaan 200D bus 2401, 3001 Leuven, Belgium}
\email{bart.vandenbussche@kuleuven.be}
\author[orcid=0000-0001-7591-1907]{Ewine van Dishoeck}
\affiliation{Max-Planck Institut f\"{u}r Extraterrestrische Physik (MPE), Giessenbachstr. 1, 85748, Garching, Germany}
\affiliation{Leiden Observatory, Leiden University, PO Box 9513, NL-2300 RA Leiden, The Netherlands}
\email{ewine@strw.leidenuniv.nl}
\author[orcid=0000-0002-8909-8782]{Stacey Alberts}
\affiliation{Space Telescope Science Institute, 3700 San Martin Drive, Baltimore MD 21218, USA}
\email{salberts@stsci.edu}
\author[orcid=0000-0002-7714-688X]{Rom\'{a}n Fernandez Aranda}
\affiliation{Centro de Astrobiolog\'{i}a (CAB), CSIC-INTA, Ctra. de Ajalvir km 4, Torrej\'{o}n de Ardoz, E-28850, Madrid, Spain}
\email{rfernandez@cab.inta-csic.es}
\author[orcid=0000-0001-7563-3636]{Andreea Petric}
\affiliation{Space Telescope Science Institute, 3700 San Martin Drive, Baltimore MD 21218, USA}
\email{apetric@stsci.edu}
\author[orcid=0000-0003-4702-7561]{Irene Shivaei}
\affiliation{Centro de Astrobiolog\'{i}a (CAB), CSIC-INTA, Ctra. de Ajalvir km 4, Torrej\'{o}n de Ardoz, E-28850, Madrid, Spain}
\email{ishivaei@cab.inta-csic.es}


\begin{abstract}

We present results from a survey of the Hubble Ultra-Deep Field using the Wide-Field Slitless Spectroscopic (WFSS) capability of the Mid-Infrared Instrument (MIRI) on JWST, demonstrating the capabilities of this new mode. We describe the data reduction and calibration methodology, and estimate calibration uncertainties. From our observations we obtain spectra of 47 galaxies with confirmed spectroscopic redshifts, with a maximum $z_{\mathrm{spec}}$ of 3.712. In the final sample we target in particular the 3.3~\um~Polycyclic Aromatic Hydrocarbon (PAH) feature, which has recently gathered interest as a star formation rate indicator and diagnostic for the dust grain size distribution in star forming galaxies from the local Universe to intermediate redshifts. The feature falls into the WFSS wavelength region for redshifts 0.67 to 3.1 - providing full coverage of the peak star formation ``Cosmic Noon'' era (1 $< z <$ 3) and connecting dust properties in this critical galaxy evolution period with local-Universe and low-redshift observations. Using the galaxies in our sample in this redshift regime, we test correlations identified in lower-redshift samples in the near-infrared or targeted programs in the mid-infrared, finding the WFSS spectra, even with large calibration uncertainties, show good agreement with complementary samples. the 3.3~\um~PAH luminosities follow previously established correlations with total IR luminosity and SED-derived star formation rates, confirming this feature's power as tracer of dust-obscured star formation. Our work illustrate the potential of the MIRI WFSS mode for studies of Cosmic Noon-era galaxies in particular in an observationally efficient way.


\end{abstract}



\section{Introduction}
\subsection{Dust at Cosmic Noon}\label{subsec:intro_dust}

The launch of JWST~\citep{2006SSRv..123..485G} has expanded the sample sizes of galaxies at high redshift substantially, increasing the diversity of objects detected, the range of masses and luminosities, and extending deeper into redshift space~\citep[e.g.][and many more]{2023ApJS..269...16R, gillman2024, 2024ApJ...964...71H, 2023ApJ...946L..13F, 2023MNRAS.520.3974C, 2023ApJ...954...31C, matthee2024, 2023ApJ...951L...1P, 2025ApJ...991..179P}. JWST's infrared (IR) coverage, and particularly the mid-infrared (MIR) coverage of the Mid-Infrared Instrument~\citep[MIRI;][]{2015PASP..127..584R, 2023PASP..135d8003W}, allows us to study the near- and mid-infared spectra of galaxies, to measure galaxy and supermassive black hole growth to redshifts inaccessible to past instruments and missions~\citep[e.g.][]{2025NatAs...9..155Z, vidal2026, 2026arXiv260220247P}. 

The mid-infrared spectra of star-forming galaxies are dominated by emission features from Polycyclic Aromatic Hydrocarbons (PAHs), complex carbonaceous molecules that represent the smallest (nano-particle sized), most common type of dust grains~\citep{2007ApJ...656..770S}. Transient heating from optical and ultraviolet emission from young stars causes emission from PAH molecules in distinctive bands, the brightest of which are found at 3.3, 6.2, 7.7, 8.3, 11.3, 12.7 and 17~\um~\citep{1984A&A...137L...5L, 1985ApJ...290L..25A, 2008ARA&A..46..289T}. PAHs play an important role in regulating the chemistry and ionization balance of the ISM~\citep{1985ApJ...291..722T, 2001ApJ...548L..73H}; the overall strength of PAH emission and the ratio of band fluxes provides constraints on the underlying radiation field, i.e. (dust-obscured) star formation and/or the presence of active galactic nuclei (AGN), the dust grain size distribution and ionization properties~\citep{2021MNRAS.504.5287R, rigopoulou_polycyclic_2024, 2025A&A...696A.135G}. Their use and calibration as direct tracers of obscured star formation goes back many years,  with early work using data from the \textit{Spitzer Space Telescope}~\citep[e.g.][many more]{2005ApJ...632..169L, 2016ApJ...818...60S}. With JWST this work can be extended to less luminous populations and a wider range of environments~\citep{rigopoulou_polycyclic_2024, 2026ApJ...997...20G, alberts2026}. 

Different observing modes on JWST have been used to advance our understanding of PAH emission in star forming galaxies beyond the local Universe. \citet{lyu2025} used NIRCam Wide-Field Slitless Spectroscopy (WFSS) from the First Reionization Epoch Spectroscopically Complete Observations survey~\citep[FRESCO; ][]{fresco2023} to probe around 200 galaxies from z$\sim$0.2--0.5. They identify 3.3~\um~PAH emission in $\sim$45\% of galaxies, providing the first statistical sample in galaxies not limited to Luminous or Ultra-Luminous Infrared Galaxies, i.e. probing into populations with log $L_{\mathrm{IR}}/L_{\odot} \ll 11$. The correlation between $L_{\mathrm 3.3}$ and $L_{\mathrm{IR}}$ is consistent with that identified in local galaxies, validating the use of this feature as a star formation rate indicator. Going beyond the low-$z$ regime, SED fitting provides an effective way of identifying PAH emission in large imaging surveys with JWST's Mid-Infrared Instrument~\citep[MIRI; ][]{2015PASP..127..584R, 2023PASP..135d8003W}. \citet{2023ApJ...946L..40L} produced predicted MIRI colour-colour tracks to identify PAH-emitting galaxies at intermediate redshifts. Using MIRI imaging data,~\citet{shivaei2024} derived PAH emission characteristics via SED fitting in a sample of z $\sim$ 0.7--2 galaxies from the SMILES survey~\citep{smiles2024}, focusing particularly on the fraction of dust in PAHs ($q_{\mathrm {PAH}}$), its correlation with stellar mass and evolution with redshift and metallicity. 

The study of dust via the mid-infrared PAH features highlights the strong synergy between JWST and ALMA, with the latter probing the dust continuum emission as well as molecular gas properties to high redshifts. PAHs are known to be tightly correlated with molecular gas over a wide range of luminosities~\citep{2013ApJ...772...92P}, and observations with JWST and ALMA have confirmed this finding~\citep{2023ApJ...944L...9L}. With JWST, this work can also be extended beyond the local Universe: ~\citet{shivaei2024} combine data from the ALMA Spectroscopic Survey in the Hubble Ultra-Deep Field~\citep[ASPECS;][]{Walter2016, Dunlop2017} and JWST Systematic Mid-infrared Instrument Legacy Extragalactic Survey~\citep[SMILES; ][]{2024ApJ...975...83R, smiles2024} to examine the correlation between (7.7~\um) PAH and CO luminosities for $1 < z < 3$ galaxies.

The 3.3~\um~PAH feature has gathered significant interest due to its diagnostic power for grain size distribution, which plays an important role in interstellar medium (ISM) heating and cooling~\citep{lai2023, 2023ApJ...942L..37A}. Poorly accessible with previous missions, the feature was only detected (spectroscopically) in a handful of very luminous or lensed galaxies with the \textit{Spitzer Space Telescope} or \textit{AKARI} mission beyond the local Universe~\citep{2009ApJ...703..270S, 2009ApJ...698.1273S, 2020NatAs...4..339L, 2026PASJ...78..454K}. JWST's broad infrared coverage and deep spectroscopic sensitivity is particularly well suited to studies of the 3.3~\um~feature, allowing it to be detected (at least theoretically) from the local Universe to $z \sim$7.4 with its complement of observing modes. Of particular interest for this work is the study of this feature in the $z \sim$ 1--3 Cosmic Noon era, where dust-obscured star formation is known to dominate the star formation rate density budget~\citep{2014ARA&A..52..415M, 2017MNRAS.466..861D, 2021ApJ...909..165Z, 2023Natur.618..708S}. 

Particularly relevant to our work is the study of the 3.3~\um~PAH emission by~\citet{mckinney2026} using the MIRI Low-Resolution Spectrometer~\citep[LRS; ][]{2015PASP..127..623K} in a sample of 37 Spitzer-selected ULIRGS (log $L_{\mathrm{IR}}/L_{\odot} > 11.5$) in the range $z = 0.65 - 2.46$.~\citet{mckinney2026} extend the work of~\citet{lyu2025} in redshift and luminosity space, confirming the correlations between $L_{\mathrm 3.3}$ and SFR, $L_{\mathrm{IR}}$. The JWST Cycle 3 program PAHSPECS (PI: I. Shivaei) targets a sample of 5 $z \sim$ 1 ALMA-selected galaxies with the MIRI Medium Resolution Spectrometer, to connect the properties of the PAH bands and their ratios to those measured in the local Universe. \citet{2026arXiv260618230L} perform an analysis of the integrated PAH emission properties, finding that different ISM conditions shape the PAH emission in this redshift regime than in local starburst galaxies. \citet{2026arXiv260618244D} use the integral field spectroscopy capability of the MRS to perform a spatially resolved analysis. By constructing spatial maps of PAH ratios, they identify radial gradients in dust grain size distribution, again highlighting differences with observations in the local Universe. These studies illustrate the diagnostic power of mid-IR PAH features for ISM physics in Cosmic Noon galaxies. In this paper we build on these projects, demonstrating the capabilities of the newly available MIRI Wide-Field Slitless Spectroscopy mode~\citep{petric2026} in the study of the 3.3~\um~PAH feature in Cosmic Noon galaxies.



\subsection{Technical capabilities and limitations for WFSS with the MIRI Double Prism}

MIRI's WFSS capability leverages the shared detector real estate from the MIRI Imager and LRS modes~\citep{2015PASP..127..623K, 2024A&A...689A...5D}. LRS slit spectroscopy is taken by placing the target in the slit, and any sources serendipitously located in the imager or other illuminated portions of the field are also dispersed. Therefore, whenever the LRS slit spectroscopic mode is operated, the imager simultaneously operates as a MIRI Wide Field Slitless Spectroscopy mode, though this mode was not supported in the early cycles of JWST operations. Note that the LRS can be operated in a different slitless mode since the start of the mission; this mode is specifically designed for single-object slitless time-series spectroscopy in the SLITLESSPRISM subarray, which measures just 416 $\times$ 72 pixels. In this work we build on the experience with these modes and the publicly available calibration reference files to test the scientific capability of a WFSS mode with MIRI. This mode was formally offered for the first time in JWST Cycle 5 (starting July 2026); but wide-field dispersed spectra are present in the MAST archive in LRS slit observations from Cycles 1--4, when sources are present in the Imager field. For the present work, we targeted arguably the best-known extragalactic field - the Hubble Ultra Deep Field (HUDF). 

The optical design of the LRS, powered by a ZnS/Ge double prism (denoted P750L in MIRI's filter wheel), provides instantaneous spectral coverage from $\sim$5 to 14~\um. Note that while the dispersion goes to 14~\um, the prism throughput drops steeply beyond 10~\um, and the 12--14~\um~range is only scientifically usable in the highest SNR spectra; sometimes  12~\um~is quoted as the red limit of the LRS/WFSS spectral range, though theoretically it extends to 14~\um. The spectral resolving power $R$ increases quasi-linearly from $\sim$40 to $\sim$160 from 5 to 12~\um~(reported in~\citet{2015PASP..127..623K}, and see also~\citet{xuan2024} for an in-flight empirical derivation). The 5-14~\um~spectra are dispersed over approximately 400 px, vertically along detector columns, with a strongly non-linear dispersion profile (see Fig.~\ref{fig:dispersion_comp}). The dispersion relation is known to vary with field location due to optical distortion effects. High-quality wavelength calibration reference files are currently available for spectra in the slit and at the nominal pointing location in the SLITLESSPRISM subarray; the reported accuracy is approximately 20 nm in the ``core'' wavelength range ($\sim$6-12~\um), with higher residuals nearer the band edges.

In the MIRI wavelength range, the JWST background becomes increasingly dominated by the thermal emission from in-field zodiacal dust (4 to 15~\um); and by the thermal self-emission beyond 15~\um~(no longer falling within MIRI's double prism bandpass). The broad passband of the double prism results in a high level of background light reaching the detector, reducing the sensitivity of the slitless LRS mode by an order of magnitude compared with the performance when using the slit\footnote{See \href{https://jwst-docs.stsci.edu/jwst-mid-infrared-instrument/miri-performance/miri-sensitivity}{JWST documentation} for the latest sensitivity curves}.

\section{Observations and data reduction}

The data presented in this paper were obtained as part of the MIRI Guaranteed Time Observations program, Program ID 4533 (PI: G. \"{O}stlin), targeting a portion of the Hubble Ultra Deep Field~\citep[HUDF;][]{2006AJ....132.1729B}.  The HUDF has tremendous legacy value, having been observed with most major space- and ground-based observatories across the electromagnetic spectrum. Particularly relevant for these observations is the extensive coverage in recent years with JWST: the JWST Advanced Deep Extragalactic Survey~\citep[JADES;][]{2026ApJS..283....6E, 2026arXiv260115956R} and the SMILES surveys~\citep{2024ApJ...975...83R, smiles2024}, and in particular the deepest-yet mid-infrared observations with MIRI F560W in the MIRI Deep Imaging Survey~\citep[MIDIS; ][]{midis2025}, recently further extended to the F770W and F1000W filters. In addition, the targeted field also overlaps with deep ALMA data from the ASPECS survey~\citep{Walter2016, Dunlop2017, Aravena2020}. As such it contains a sample of well-characterized dusty galaxies, providing an ideal test target field for these mid-infrared observations. 

A key design idea in the program was to obtain a slit and slitless spectrum of a known source, to provide a benchmark for the calibration quality in the WFSS data, though as the APT template for MIRI WFSS did not exist at that time, the observations were entirely set up as MIRI LRS slit observations. The galaxy chosen to provide these benchmark spectra is the ALMA-detected source 1mm.C10~\citep[hereafter `ALMA-C10';][]{Aravena2020} - its coordinates were entered into APT as `science target'. ALMA-C10 is included in the ASPECS survey main sample~\citep{Aravena2020}, with a spectroscopic redshift of $z = $ 1.997, measured independently from optical and CO spectroscopy~\citep{2017A&A...608A...2I, boogaard2020}; \citet{Boogaard2024} associate it with an X-ray source in the Chandra-Deep Field South catalog from \citet{2017ApJS..228....2L}. \citet{Boogaard2024} report a stellar mass of 6.31 $\pm$ 1.45 $\times 10^{10} M_{\odot}$ and effective radius $R_e$ of 2.934 $\pm$ 0.147 kpc (as measured in MIRI F560W). \citet{Aravena2020} report a stellar mass of 12.6 $\pm$ 1.3 $\times 10^{10} M_{\odot}$ and total infrared luminosity ($L_{\mathrm{IR}}$) of 6.6$^{+3.9}_{-2.3} \times 10^{11} L_{\odot}$. Imaging (see Fig.~\ref{fig:slit_placement}) shows the galaxy to be spatially extended with at least one resolved spiral arm-like feature. Kron fluxes from the SMILES survey in F560W, F770W, F1000W and F1280W are 29, 19, 27 and 34~$\mu$Jy, respectively~\citep{smiles2024}. 

In the first observation, ALMA-C10 was placed into the LRS slit, and observed with a 3-row slit-stepped mosaic (20\% row overlap). At each mosaic position, the standard along-slit-nod dither pattern is executed, resulting in a set of 6 exposures ($t_{\exp} =$ 18 min per exposure). Prior to the mosaic, a 166-s verification image was taken with the F560W filter; this image acts as source locator for the slitless spectra in the imager field. No target acquisition was performed. 

A second observation was configured with an offset in both translation (16~arcsec in x, 8~arcsec in y) and position angle ($\Delta$PA=5$^{\circ}$) compared with observation 1. This places the ALMA-C10 galaxy in the imager field of view, and provides a position angle (PA) rotation of 5 degrees to the dispersed field to separate spectra that may overlap a the first PA (and vice versa)\footnote{This was implemented in the APT proposal via the PA Offset Link special requirement, for which we provided a range of 5-10 degrees.}. This observation performed a 4-point ``Mapping'' pattern, stepping 7~arcsec and 10~arcsec in spatial (x) and spectral (y) directions, respectively ($t_{\exp} =$ 31.5 min per exposure). A verification image was taken with the same configuration as that of Observation 1 for source locations. 

This strategy provides coverage of the HUDF field at two position angles and with small and larger spatial offsets; and a well-known and well-characterized galaxy in both the slit and the imager field, allowing a direct comparison of slit and slitless spectra. The total on-sky exposure time, combining the 2 observations, is 3.9 hours. 

The slit spectra of ALMA-C10 were largely reduced and calibrated using the JWST calibration pipeline (v1.17.1), with few modifications. As no calibration pipeline exists for dispersed spectra in the MIRI imager field, the data reduction and calibration for this portion of the dataset was performed entirely using custom code; though we note our custom steps follow the same overall workflow as the JWST calibration pipeline for other WFSS modes. With both in-slit and slitless spectra available for the reference target, ALMA-C10, we are able to perform a basic test on calibration accuracy for the slitless spectra. We describe the different calibration steps in the sections below.

\subsection{Detector-level data reduction: From counts to slopes}

The basic detector calibrations for our data were performed using the JWST \texttt{calwebb\_detector1} calibration pipeline (v1.17.1, CRDS context 1321). No custom modifications were required. The \texttt{\_rate.fits} files were used as starting point for the remainder of the reduction.

\subsection{Pointing reconstruction and astrometry}\label{sec:pointing}

\begin{figure}
    \centering
    \includegraphics[width=\textwidth]{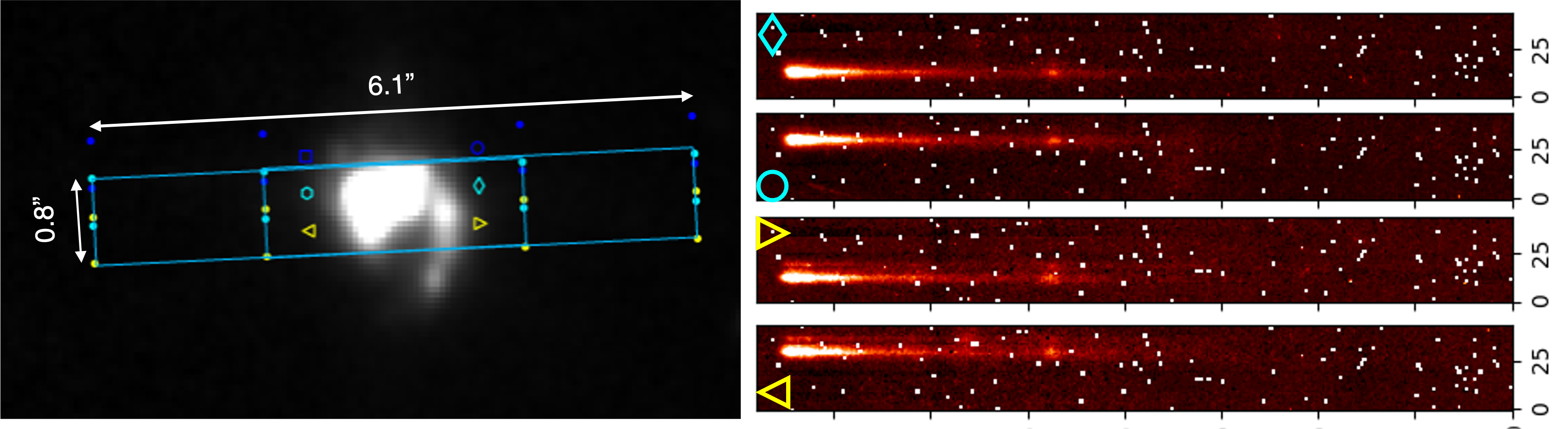}
    \caption{Left: MIRI F560W image of target galaxy ALMA-C10 from the MIDIS survey~\citep{midis2025}, showing the positioning of the LRS slit in Obs 1 of program 4533 overlaid. The mosaic contained 3 pointings along the dispersion direction, indicated in blue, cyan and yellow, respectively. An along-slit-nod pattern executed at each position, resulting in 6 exposures. The filled round symbols indicate the slit vertices in each pointing, and the open symbols indicate the slit center location. The placement has been corrected for the systematic FGS-MIRI pointing offset that resulted in the galaxy being largely missed in the first mosaic pointing (exposures 1 and 2). Right: 2D spectral images of the galaxy at 4 of the 6 pointing locations (exposures 3, 4, 5 and 6). The background has been subtracted, and the images are rotated by 90 degrees for visualization purposes; the dispersion direction of LRS is vertical. Each spectral image is accompanied by a cyan or yellow symbol indicating the pointing at which the spectrum was obtained. In exposures 5 and 6 (yellow triangle symbols), the spectrum of the spiral arm is clearly resolved.}
    \label{fig:slit_placement}
\end{figure}

First inspection of the Observation 1 exposure sequence revealed that the ALMA-C10 galaxy was not visible in the LRS slit in verification image, despite being visible at high SNR in the direct image for Observation 2 with a similar exposure time. Further investigation revealed that the observation was affected by a systematic calibration offset of approx. 0.15-0.2~arcsec in the telescope v3 axis discovered in autumn 2024 between the JWST Fine Guidance Sensor (FGS) and the MIRI instrument~\citep{sohn2024}\footnote{This issue was resolved in September 2025}. As a result, the first 2 of 6 exposures do not contain any detectable source flux. As ALMA-C10 is clearly detected in the 2nd and 3rd mosaic rows, we chose not to repeat the observations. In Fig.~\ref{fig:slit_placement} we show the reconstructed (i.e. corrected for the offset) positioning of the slit mosaic on the galaxy, with background-subtracted 2D spectral images for the 4 well-positioned exposures alongside. 

For the dispersed spectra in the Imager field, we use the standard \texttt{calwebb\_detector1} reduction for initial detector calibration steps. Beyond this, no pipeline support was available for this portion of the data, at the time the data were taken and reduced. We describe in the following sections the calibration steps performed to extract final, flux-calibrated spectra for the slitless dispersed data:

\begin{itemize}
\item{Reconstruction of spatial offsets between dithered exposures}
\item{Source catalog construction}
\item{Background subtraction}

\item{2D Source extraction}
\item{Source extraction, wavelength and flux calibration}
\item{Combining spectra from multiple exposures}
\end{itemize}

\subsection{Reconstruction spatial offsets between dithered exposures}\label{subsec:reduction_offsets}

To be able to associate sources between the different exposures and observations, the offsets must be known to high accuracy. As  World Coordinate System (WCS) information is only available in the LRS slit region for the WFSS data, we cannot use sky coordinates in the imager field to associate sources between exposures. We therefore derive the offsets in pixel coordinates. The first WFSS exposure and the direct image, from which source positions will be measured, form a matched pair with no offsets apart from the small ($<$ 1 px) boresight offset between the F560W and F770W filters (the double prism pointing position is calibrated against the F770W filter).  Subsequent dispersed exposures will have a spatial and/or spectral offset with respect to this pair.  

To identify the offsets, we run the exposures through the \texttt{assign\_wcs} step in the \texttt{calwebb\_spec2} pipeline, and use the ra\_ref and dec\_ref attributes of each exposure to determine its slit centre sky coordinates. Using the WCS information of the direct image, we convert these slit centre coordinates back to (x, y) locations on the detector, and compute (dx, dy) offsets for each image with respect to the direct image.

\subsection{Source catalog}\label{subsec:catalog}

We use the direct images taken in F560W for each observation to create a catalog of sources. As only 1 dither is available for these images, they have relatively poor cosmetics. We calibrate the images up to the flux calibration \texttt{photom} step in the JWST pipeline. We use standard \texttt{photutils} source finding tools to identify all sources with SNR $\geq$ 5.0. We cross-matched the sky coordinate of each source with the source catalogs from the MIDIS and JADES surveys to obtain redshift information and photometry. 

\begin{figure}
    \centering
    \includegraphics[width=0.9\linewidth]{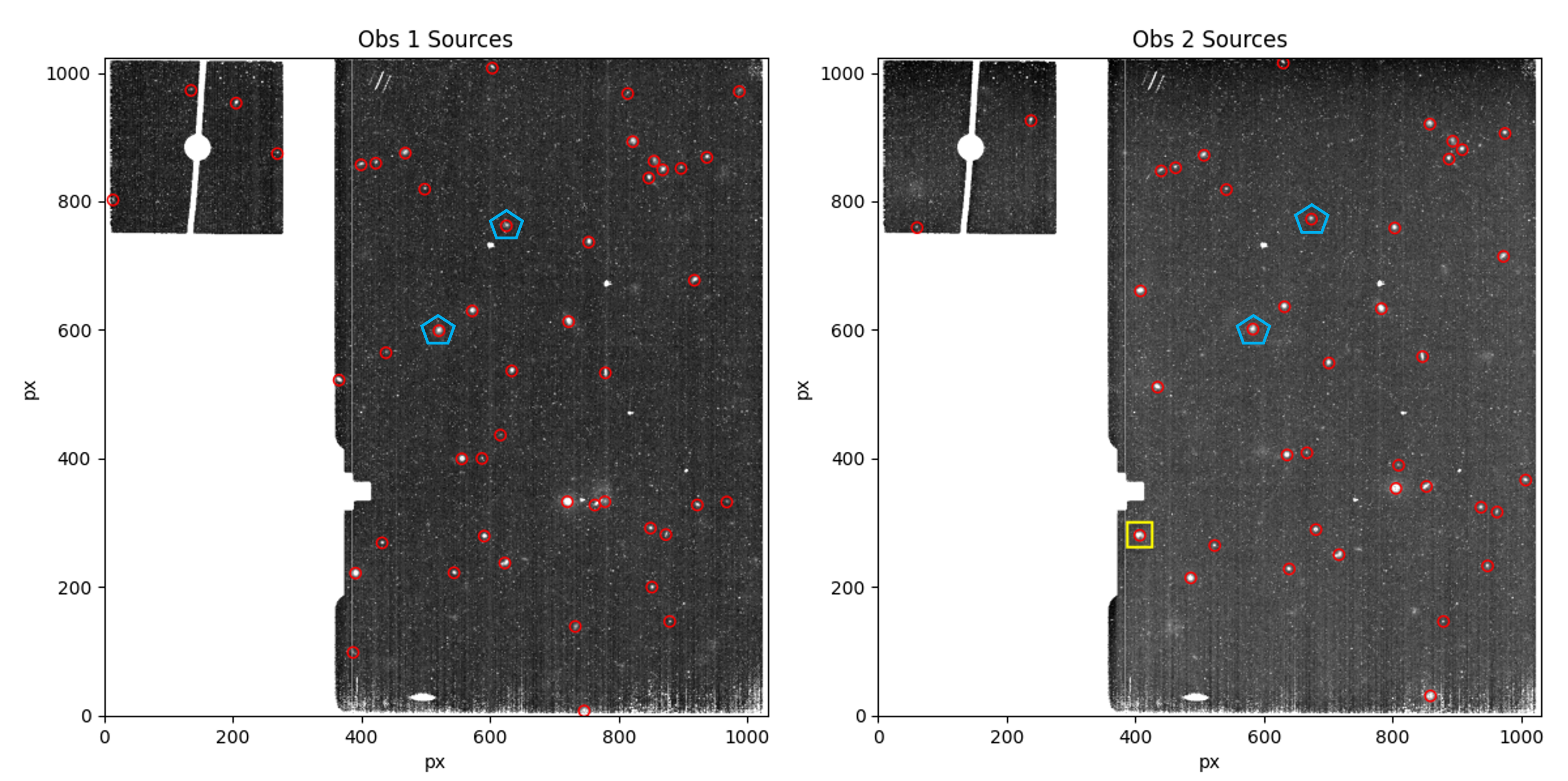}
    \caption{Direct images of the 2 observations, taken in the F560W filter with an exposure time of 166 s; the sources detected with SNR $\geq$ 5 are circled in red. In the Observation 2 image (R), reference target ALMA-C10 is marked in the yellow box. In the Observation 1 image, ALMA-C10 is located in the LRS slit, which is not visible in this view. Note that these are early source detection images; not all produced usable spectra. The sources very close to the detector edges or the Lyot mask were excluded from the sample. The two galaxies marked with the blue polygons are those described in Section~\ref{subsec:pahspecs_comp}. Direct images for LRS slit observations are taken in a single exposure without dithering, so the image cosmetics are relatively poor.}
    \label{fig:direct_images}
\end{figure}

We find 47 sources in observation 1, 39 sources in observation 2. 34 sources are detected in both observations. Accounting for the overlap between the observation 1 and observation 2 fields, we detect 51 unique sources. On inspection of the dispersed data (see also below), 4 of these were removed from the sample because they were too close to the detector edge or the Lyot mask, or too blended with nearby sources to yield useful spectra, or spurious detections  - giving a final sample size of 47. Note that we ignore any sources in the regions of the detector occupied by the 4-quadrant phase masks as these regions cannot be calibrated in the same way as the Imager field. Additional sources can likely be identified using pre-existing catalogs from deeper surveys; however given the relatively shallow nature of our observations and the lack of full calibration, we focus on the brightest, high-SNR sources. Fig.~\ref{fig:direct_images} shows the direct images with detected sources. 

\subsection{Background subtraction}\label{subsec:background}

\begin{figure}
    \centering
    \includegraphics[width=\linewidth]{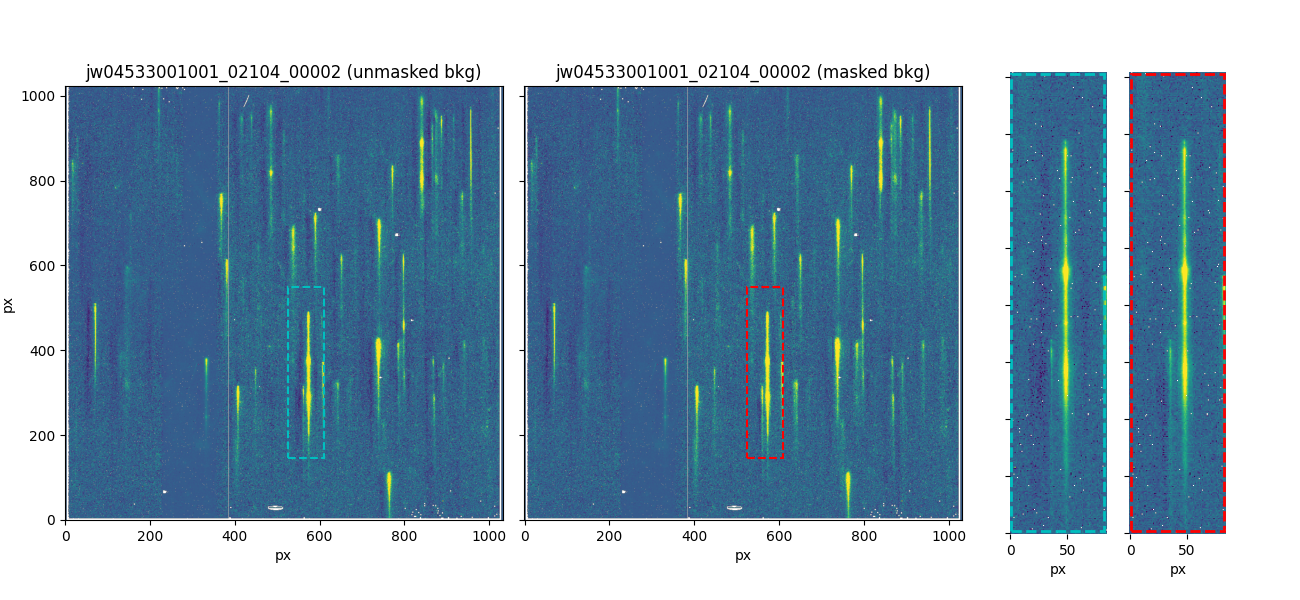}
    \caption{A background-subtracted exposure from the first observation in our program, illustrating the difference between background creation methods. From left to right: (i) exposure subtracted with a master background created without source masking; negative traces are visible alongside many of the spectra. (ii) the same exposure, with the background created from all exposures with sources masked with a fixed rectangular aperture; most over-subtraction artifacts are absent or greatly reduced. Some negative trace features remain, especially near crowded or extended sources. (iii) and (iv) a zoomed-in region from each exposure showing a detailed view of the changes between methods. The detail view for the improved method (red box) still shows some residual over-subtraction in the region where sources are closely spaced.}
    \label{fig:bkg_method}
\end{figure}

The broad bandpass of the double prism transmits the full 5-14~\um~background spectrum, dispersing this quasi-uniformly across the imager field.. The background levels dominate the source spectra prior to subtraction in fields such as ours, which contains predominantly faint, compact extragalactic sources. As all exposures in this program were taken as a sequence, there is no significant change in background levels seen between them. We produce a background from all 10 dispersed exposures, using the source catalog and exposure offsets information to locate sources in the field, and masking a 10 $\times$ 400 px rectangle centred on the spectral trace. We produce a median of the (10) source-masked exposures, and subtract this from all dispersed exposures, revealing the fainter spectra of the HUDF sources. Without source masking the background subtraction results in significant regions of over-subtraction. This is illustrated in Fig.~\ref{fig:bkg_method}. The procedure can likely be optimised by taking the size of individual sources into account (e.g. using variable-sized masking regions), or by masking based on deeper imaging to account more comprehensively for fainter sources in the field.

\subsection{2D Source extraction and association}\label{subsec:2dextarct}

\begin{figure}
    \centering
    \includegraphics[width=0.9\linewidth]{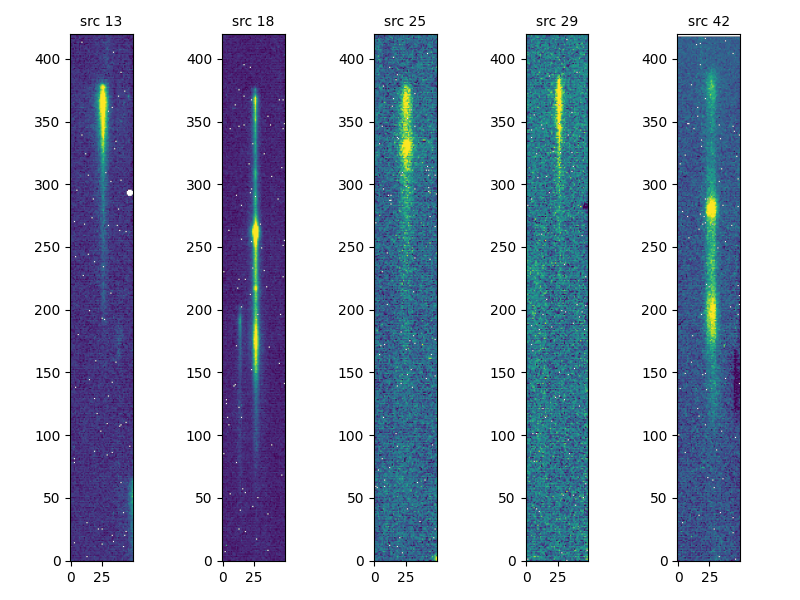}
    \caption{Example 2D spectral image cutouts of different targets in our observations. These are the cutout sizes used for further calibration and extraction.}
    \label{fig:cutouts}
\end{figure}

As the dispersion profile of the MIRI double prism changes with field location, and this variation is at this point not very well characterised, we do not co-add spectral images in this reduction. Using the offset information gathered in section~\ref{sec:pointing}, we locate each source in all exposures, create and store image cutouts measuring 50 $\times$ 420 px - dimensions known to capture the spectral trace of the LRS. The trace shape has been characterized in detail for the LRS slit and SLITLESSPRISM locations; it shows very mild curvature at the red end of the spectrum (moving across $\sim$1-2 pixels over the full range). The shape is however also variable with field position due to optical distortion. An example of a set of cutouts for a given target is shown in Fig.~\ref{fig:cutouts}.

We record in the headers of each cutout image the source information: its (x, y) location in the parent exposure as well as in the cutout itself; the source label; the spatial offset from the reference location in the direct image; and, where available, the spectroscopic redshift. 

\subsection{Spectral extraction and calibration}\label{subsec:extract1d}

Prior to spectral extraction, we perform a row-by-row subtraction of the median sigma-clipped residual background in each cutout (masking the spectral trace itself); this accounts for local under- or over-subtraction of the background. Following this adjustment, a simple boxcar extraction algorithm was used to extract spectra from the images. The extraction and wave calibration should ideally incorporate the spectral trace shape; but given the very small amount of curvature in the trace, we ignore this in our current reduction. The extraction aperture was adjusted to the size of each source, measured by collapsing the cutout in the spectral direction to visualize the location and width of the trace. On the extracted spectra we perform 2 further calibration steps:

\begin{enumerate}
    \item Wavelength calibration: we use the \texttt{specwcs} reference file currently used for LRS slit spectra (\texttt{jwst\_miri\_specwcs\_0010.fits}) to convert pixels to wavelengths;
    \item Flux calibration: we assume that the spectral response function, i.e. the conversion from DN/s to physical units as a function of wavelength for WFSS spectra is the same as that of MIRI single-object slitless spectra, and apply the slitless LRS photometric calibration to the WFSS spectra (\texttt{jwst\_miri\_photom\_0229.fits}). 
\end{enumerate}

We first test this on our ``nominal'' science target ALMA-C10, for which we have a well-calibrated LRS slit spectrum as well as the slitless spectrum taken in the imager field. We subsequently apply the strategy to all WFSS sources in the field. Both wavelength and flux calibration reference files are publicly available and commonly used in the JWST calibration pipeline; the files were applied using the same algorithm as used in the pipeline.

To complement our data and provide checks on the photometric calibration, we cross-matched our WFSS source catalog against catalogs from complementary surveys: MIDIS~\citep{midis2025} and SMILES~\citep{smiles2024, 2026arXiv260115956R}.

\subsubsection{ALMA-C10 as test case}\label{subsec:alma10}

The observational strategy of placing ALMA-C10 into the LRS slit for Obs 1, then offsetting for Obs 2 such that the galaxy is located in the imager field, provides us with high-SNR spectra for the same source, dispersed both in the slit and subsequently as a WFSS source in the imager field. This is an good test scenario for the calibration procedures for the WFSS sources. 

\begin{figure}
    \centering
    \includegraphics[width=\textwidth]{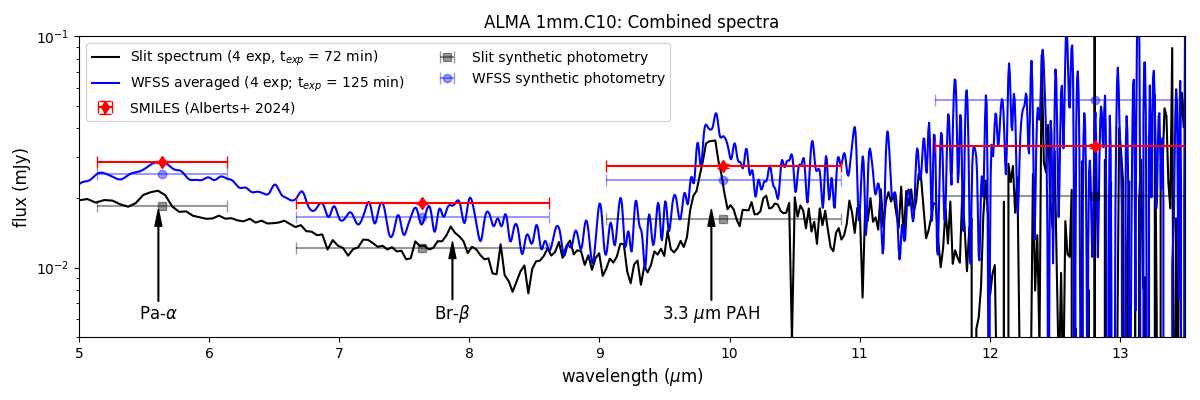}
    \caption{Comparison of the flux-calibrated summed LRS slit spectrum of galaxy ALMA-C10 (black) and the WFSS spectrum, co-added over 4 exposures (blue). Red points show photometry in filters F560W, F770W, F1000W and F1280W filters from the SMILES~\citep{smiles2024} survey. Blue and black points show synthetic photometry computed from the WFSS and slit spectra, respectively. Positions of key features detected in the spectra (Pa-$\alpha$, Br-$\beta$ and the 3.3~\um~PAH feature) are indicated.}
    \label{fig:alma10_wfss_vs_slit}
\end{figure}

For the slit spectra, we process the data through all three stages of the JWST calibration pipeline. We do not combine all 4 good mosaicked exposures; for the final 3rd pipeline stage we produce association files for the nodded pair at each mosaic pointing (i.e. exposures (3, 4) and exposures (5, 6), as shown in Fig.~\ref{fig:slit_placement}). Then we sum the 2 final spectra to obtain an ``integrated'' spectrum of the galaxy, from slit spectroscopy; though we note that the mosaic does not cover the full extent of the galaxy. For the slit spectra, each nodded pair represents an exposure time of 36 min. The WFSS exposures had an exposure time of $\sim$30 min each, i.e. a total WFSS exposure time of $\sim$2.1 hrs.  In Fig.~\ref{fig:alma10_wfss_vs_slit} we show the summed slit spectrum of ALMA-C10, together with the co-added WFSS spectrum, photometry from the SMILES survey~\citep{smiles2024}. 

Spectral extraction was performed with the same spatial aperture size. We find good qualitative agreement between the datasets, considering the lack of dedicated calibration files and calibration pipeline for the WFSS data. With a known spectroscopic redshift of 1.998, we observe the 3.3~\um~PAH feature, and a weaker detection of Pa-$\alpha$ ($\lambda_{\mathrm{rest}}$ = 1.87~\um) and Br-$\beta$ ($\lambda_{\mathrm{rest}}$ = 2.63~\um). The slit spectrum shows a better signal-to-noise ratio, as expected due its greater sensitivity; the WFSS spectrum however provides a better match to the photometry from imaging surveys, consistent with the limited spatial coverage provided by the slit, even when using a mosaicking approach. The shape of the continuum shows no significant difference between slit and WFSS spectra in the region below 11.5~\um~(the spectrum is too noisy at longer wavelengths). The spectral features are broader with WFSS compared to the slit spectrum. A narrow slit naturally increases the resolving power of a spectrometer (all else being equal, as is the case here). An additional likely explanation is the slitless mode capturing the full spatial extent of the galaxy, which is substantially larger than the slit width along the dispersion axis. As shown in Fig.~\ref{fig:slit_placement}, the 2 well-positioned mosaic rows fail to capture the full extent of the galaxy. The slit spectrum was created by co-adding these 4 exposures, each representing a portion of the galaxy spectrum convolved with the slit width, not the full size of the galaxy. A portion of the arm feature visible in the image is not covered by the slit placement, extending redwards in the dispersion direction; this may explain the red tail seen in the 3.3~\um~PAH feature profile (perhaps combined with flux from the weaker 3.4~\um~aliphatic feature). Small wavelength offsets between the individual WFSS exposures may also contribute to some broadening. Note that the extraction aperture of the WFSS and slit spectra was chosen to encompass the full extent of the galaxy (16 px). For optimal SNR in the spectrum, a smaller aperture is more appropriate. The WFSS spectrum in particular appears affected by some fixed-pattern noise. 

We use the \texttt{stpsf} and \texttt{synphot} packages~\citep{2025zndo..15747364P, 2018ascl.soft11001S} to compute synthetic photometry from the two spectra in the F560W, F770W, F1000W and F1280W filters. These points are also shown in Fig.~\ref{fig:alma10_wfss_vs_slit}. The ratios between WFSS and SMILES photometry in these filters is 0.89, 0.87, 0.87 and 1.57, respectively; between slit and SMILES photometry: 0.65, 0.64, 0.60 and 0.61, respectively. The WFSS spectrum is higher by 37-45\% than the slit spectrum over the 5-10~\um~range. While the spectra are dominated by noise past $\sim$11.5~\um~and the F1280W values are likely not physically representative, the three shorter-wave filters show very consistent flux ratios. This suggests the differences are largely due to differences in measurement apertures, slit losses or aperture corrections for the slit spectrum\footnote{We note that the spectrophotometric calibration accuracy for LRS fixed-slit spectra is reported as $\leq$2\% around 6-7~\um, increasing to 3-8\% at red wavelengths; for LRS slitless spectra larger uncertainties are reported for faint standards, rising to 20-30\% at the reddest wavelengths~\citep{2016jdox.rept......}}. 

This test validates that the wavelength calibration used by the JWST pipeline for fixed-slit spectroscopy with MIRI LRS applies well to WFSS spectra taken in the same region of the detector. In addition, the photon conversion efficiency of the LRS single-object slitless spectroscopy is valid for WFSS spectra, matching photometry to within $\leq$15\%; this difference may be larger for sources near the edges of the field where the calibration accuracy is likely to be worse. These calibrations will be improved once full pipeline support is available for the mode.

\subsubsection{Extraction and calibration of WFSS sources}\label{subsec:wfss_extraction}

We produce spectra from the WFSS exposures from all other detected galaxies in the direct images using the same methods as described for ALMA-C10 above. For the full set of sources, there are additional considerations:

\begin{itemize}
    \item For galaxies that are located near upper and lower detector edges, only incomplete spectra are available. Parts of the cutout that are off the edge of the detector are flagged in the extracted spectra, to exclude from subsequent calibration steps; 
    \item For galaxies whose spectra overlap with those of other nearby sources, we flag the affected portions of the spectra to exclude them from the co-added spectrum (based on visual inspection of the 2D spectral image cutouts and the extracted spectra). For the majority of sources that are blended in one observation, an isolated spectrum is available from the other due to the position angle offset between observations;
    \item Galaxies whose spectrum overlaps fully with a close neighbour were removed from the sample;
    \item We further mask spectral regions affected by poor background subtraction, or other data quality issues;
    \item As the dispersion is known to vary with field location, we expect the wavelength calibration accuracy to deteriorate with the sources' distance from the slit;
    \item The flux calibration is defined as a function of wavelength, so inaccuracies in wavelength calibration also affect the accuracy of the flux calibration. 
\end{itemize}

\begin{figure}
    \centering
    \includegraphics[width=\textwidth]{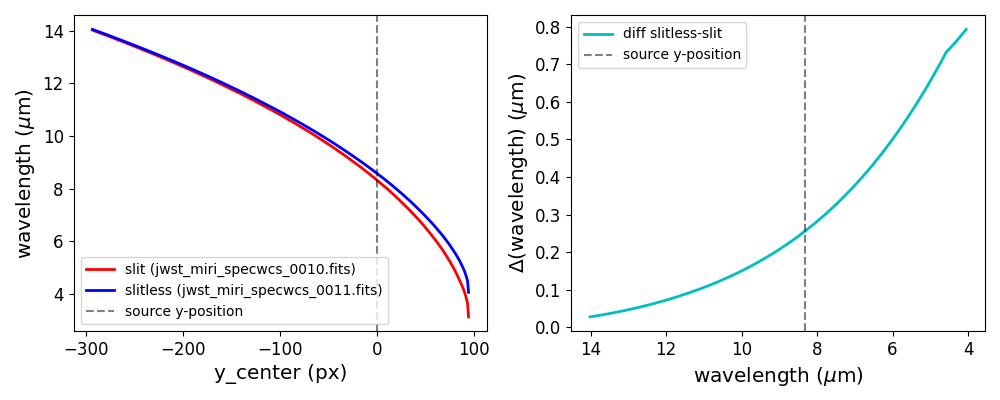}
    \caption{Comparison of the MIRI LRS dispersion profiles at the two well-calibrated locations: in the slit centre and at the nominal position in the SLITLESSPRISM subarray. Overplotted is the difference between the two profiles, illustrating the potential calibration errors in the WFSS dataset, which are all calibrated with the slit calibration reference file.}
    \label{fig:dispersion_comp}
\end{figure}

We use the publicly available calibration reference files to examine the impact of uncorrected field distortion on the wavelength and flux calibration. In Fig.~\ref{fig:dispersion_comp}, we show the dispersion relations at the two well-characterized detector locations - the slit centre position and the nominal pointing position in the SLITLESSPRISM subarray - as well as the difference between them. The coordinates of these points on the detector are (x, y) = (326, 300) and (38, 829), rounded to the nearest pixel, respectively. The figure shows that the wavelength shifts are strongly non-linear, with the largest differences at the blue end of the wavelength range, and can measure up to $\sim$0.7~\um~around 5~\um.

\begin{figure}
    \centering
    \includegraphics[width=\linewidth]{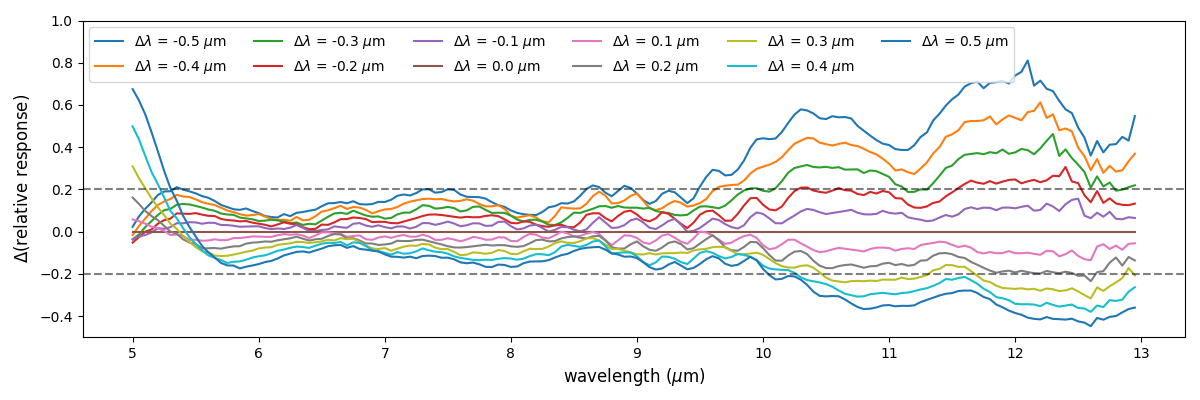}
    \caption{Estimation of the spectrophotometric calibration error from uncorrected field distortion. The plot shows the change in the relative spectrophotometric response factors, computed from the LRS slitless photometric calibration reference file, for wavelength shifts of -0.5 to +0.5~\um, over the full WFSS wavelength range.}
    \label{fig:dphotom_test}
\end{figure}

Using the photometric calibration reference file for the LRS slitless mode - the closest equivalent for the WFSS mode in terms of spectral response - we can estimate the flux calibration uncertainty as a function of wavelength shift, versus wavelength. In Fig.~\ref{fig:dphotom_test} we apply wavelength shifts of -0.5 to +0.5~\um~to the wavelength axis in the calibration reference file, and compute the resulting difference in the relative response factors. Between $\sim$5.5 and 9.5~\um, even relatively large wavelength offsets only cause flux errors within $\pm$20\%. Below 5.5~\um~and above 9.5~\um, errors become larger, up to 80\% for the most extreme cases. However, examining Fig.~\ref{fig:dispersion_comp}, we note that even for relatively far-separated locations on the detector ($\Delta$x = 288 px, $\Delta$y = 529 px for the slit and slitless locations), the wavelength shift at 10~\um~is $\sim$0.15~\um; at 12~\um $<$ 0.1~\um. It is therefore reasonable to estimate a flux calibration uncertainty in the range of 10--20\% due to uncorrected field distortion between 5.5 and 12~\um; only in the range of 5 - 5.5~\um~are uncertainties likely larger due to the strong non-linearity of the dispersion profile in this range. 

We did not attempt to re-calibrate the spectra with the preliminary dedicated WFSS calibration reference files that are now available in the JWST calibration pipeline, as these only became available late in our analysis work. 

For each galaxy, a spatial profile was constructed by summing over all rows in the cutout images. Based on this profile, the extraction aperture was defined to capture the full signal from the galaxy. Each individual spectrum was wavelength and flux calibrated using the same reference files.

\subsubsection{Final WFSS spectra}\label{subsec:wfss_final_spectra}

\begin{figure}
\centering
\includegraphics[width=15cm]{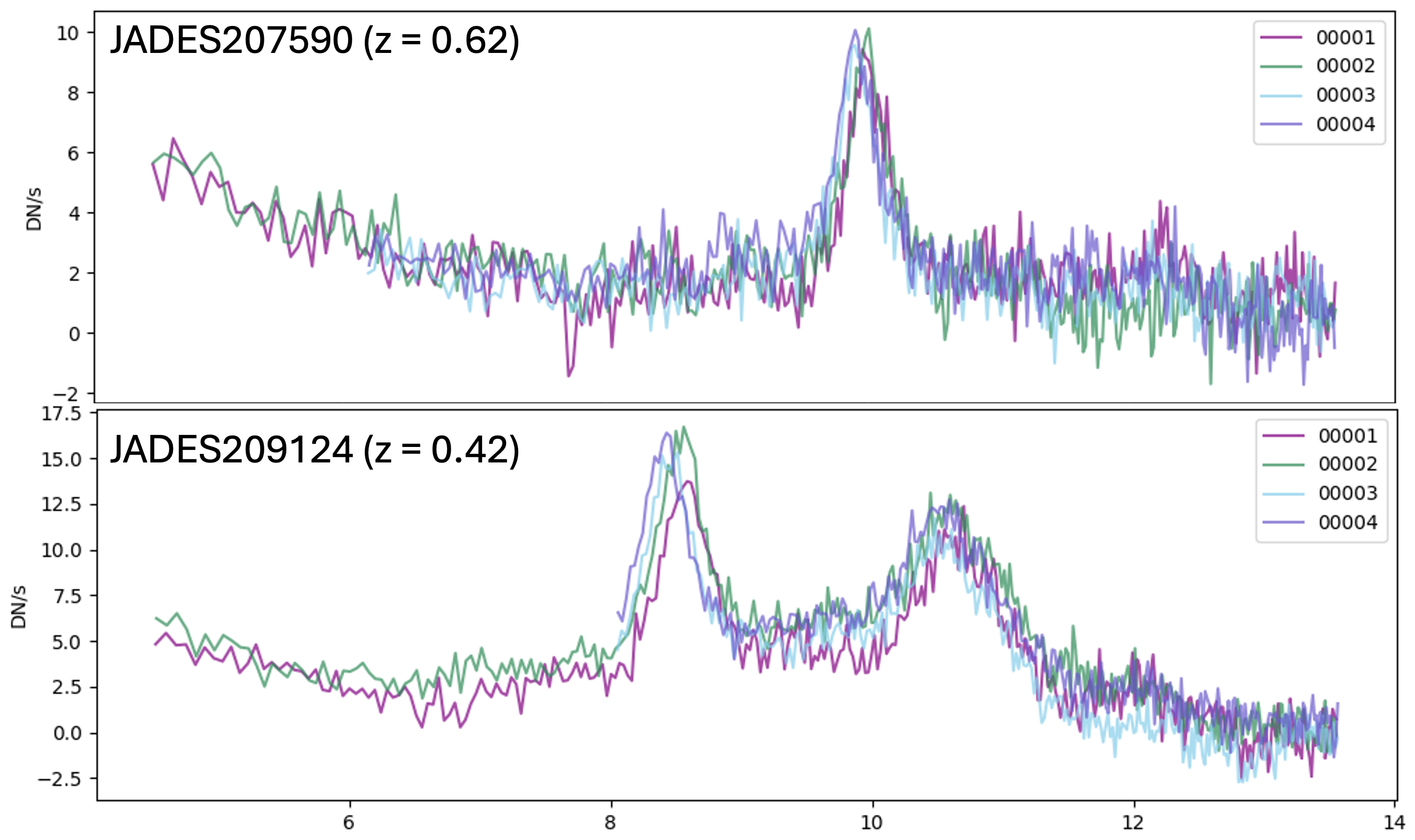}
\caption{Example spectra from individual exposures in Obs 2 showing the dispersion change between the exposures, which were taken in a 2 $\times$ 2 pattern on the detector. The offset is noticeably pair-wise, with exposures 1 and 2, separated along the x-axis but at the same y-axis location, closely matched, but offset from exposures 3 and 4. Note both sources have only partial coverage in exposures 3 and 4 due to their location near the detector top edge. JADES source IDs and spectroscopic redshifts are shown in each panel. Spectra are shown prior to flux calibration (y-axis is in units of DN/s). }\label{fig:obs2_offsets}

\end{figure}

The final stage of the WFSS calibration process is to combine the spectra from individual exposures into a final spectrum. Because the dispersion relation is known to be variable with spatial position due to field distortion, the individual spectra of each target were first visually inspected for calibration offsets. Obs 1 used a mosaic pattern with small offsets; the maximum spatial and spectral offsets between exposures were $\sim$1.8" and $\sim$0.9", respectively. These spectra are closely co-located on the array and show no measurable changes in dispersion. For Obs 2 however a larger 2 $\times$ 2 mapping pattern was performed with 7" spatial and 10" spectral steps (to provide better separation of spectral traces), as an alternative to the mosaicking approach of Obs 1. On inspection, the 4 exposures taken in Obs 2 show a distinct pattern of dispersion changes, with exposures offset in the spectral direction showing the biggest difference; the spatial shift appears to have much lesser impact. This is illustrated in Fig.~\ref{fig:obs2_offsets} for two example sources. The magnitude of the change is variable with the position of the source on the detector. The difference is most marked at the blue end of the spectra, which is expected given the steepness of the dispersion relation in this region, and for the sources closest to the detector edges. 

For some sources, these differences are small enough not to impact the final spectrum quality. For others, it may affect the measured line fits. We tested two approaches for combining spectra from Obs 2: in the first, we combined only the spectra at the same y-coordinate (along the spatial axis). As a result, galaxies that were present in both observations and at all pointing positions would have 3 resulting ``final'' spectra: Obs 1 (6 exposures), Obs 2$_{12}$ (2 exposures) and Obs 2$_{34}$ (2 exposures). In the second approach we co-added all spectra onto a common wavelength grid, regardless of small dispersion changes. We found the gain in signal-to-noise from co-adding all spectra outweighed the impact on the line profiles, and perform the analysis in the following sections on this single, co-added spectrum. Prior to co-adding the final spectra, we performed a final cleaning step where regions of contamination or poor background subtraction, as determined from visual inspection of dispersed images, were masked. Uncertainties were derived from the standard deviation of the combined spectra. For the galaxies covered in all exposures, the final spectra thus consist of 10 individual co-added spectra. For those covered only in Observation 1, 6 spectra; for those covered only in Observation 2, 4 spectra. The final spectra of our full sample are shown in Fig.~\ref{fig:all_sources_1}. 



\subsubsection{Calibration comparison with MIRI MRS}\label{subsec:pahspecs_comp}

Two galaxies in our sample have recently been observed with MIRI MRS, as part of the PAHSPECS program (see Section~\ref{subsec:intro_dust}). PAHSPECS targeted a sample of five ALMA-selected galaxies at z $\sim$ 1. This provides the opportunity to test the WFSS calibration against the mature wavelength and flux calibration of the MRS mode, for 2 additional sources at different locations in the field. These 2 galaxies, known in the samples of~\citet{Aravena2020} and~\citet{Boogaard2024} as 1mm.C16/3mm.06 ($z_{\mathrm{spec}}$ = 1.0952) and 3mm.11 ($z_{\mathrm{spec}}$ = 1.0965), respectively, are identified in the direct images shown in Fig.~\ref{fig:direct_images} with the blue symbols. The MRS exposure time for 1mm.C16/3mm.06 was approximately 10 ksec for the MRS Long setting, 12 ksec for MRS Medium and 3 ksec for MRS Short, taken over a 4-point dither pattern; for 3mm.11, the exposure times were 18 ksec for MRS Long, 12 ksec for MRS Medium and 3 ksec for MRS Short, with the same dither strategy. 

\begin{figure}
    \centering
    \includegraphics[width=\textwidth]{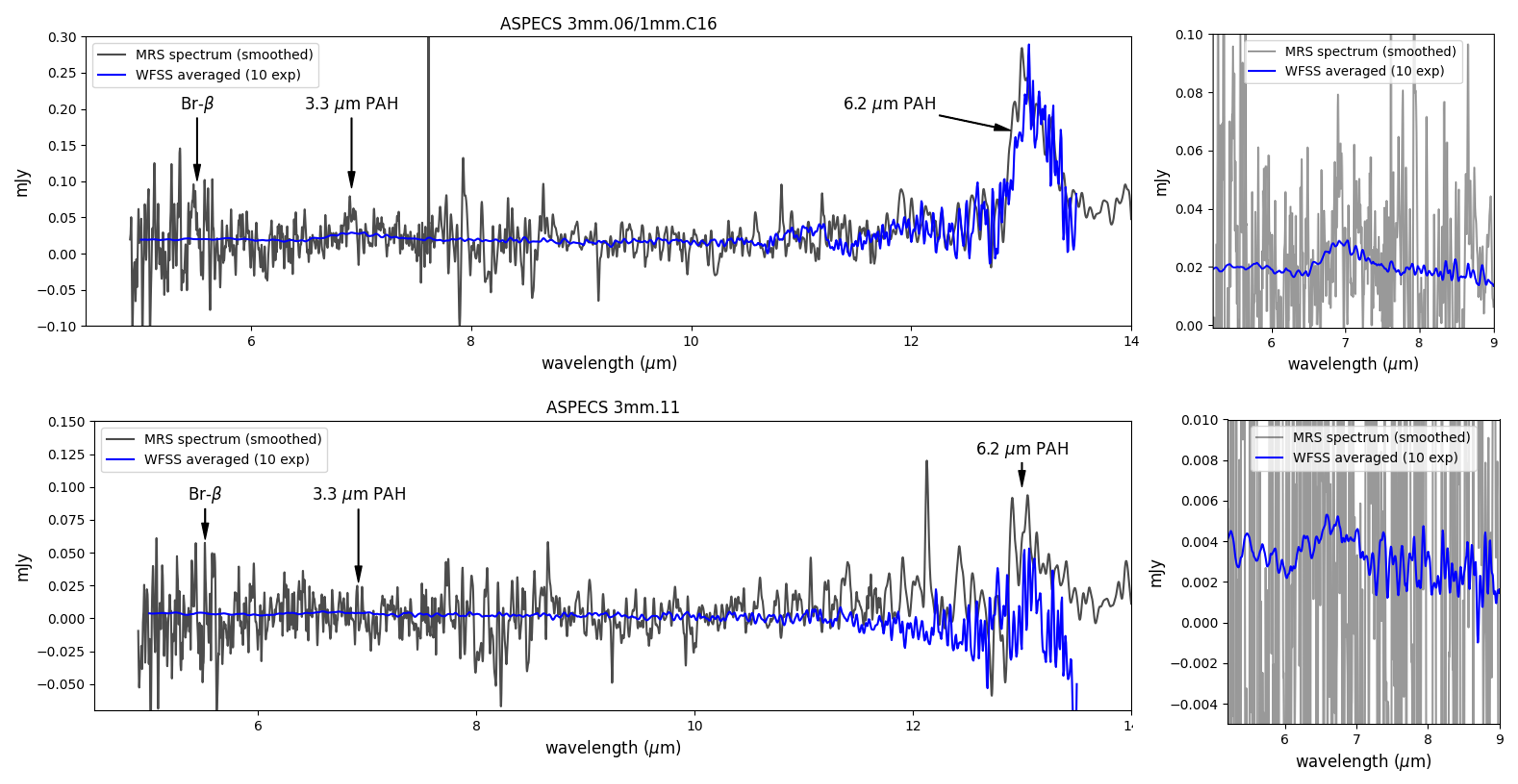}
    \caption{Comparison of the WFSS spectra from this program with the MRS spectra of the same sources from the PAHSPECS program~\citep{2026arXiv260618230L}. Key spectral features are indicated. The smaller panels on the right show the region of the 3.3~\um~PAH feature. The MRS spectra are median smoothed with a 5-px kernel for better visualization. }
    \label{fig:pahspecs_comp}
\end{figure}

The WFSS data were reduced, calibrated and extracted as described in the above sections. Both galaxies were covered in all 10 exposures, and we co-add all 10 spectra as the final reduction step. The MRS reduction and calibration is described in~\citet{2026arXiv260618230L}. Fig.~\ref{fig:pahspecs_comp} show comparisons of the MRS and WFSS spectra, with a detail view of the 3.3~\um~PAH feature. Galaxy 1mm.C16 has strong, well detected features; in the WFSS spectrum the 6.2~\um~PAH feature is shifted redwards by $\sim$ 50 nm, the 3.3~\um~PAH feature matches the MRS wavelength to around 20 nm. Galaxy 3mm.11 is fainter, and the spectra are noisier with less clear detection of even the strong 6.2~\um PAH feature. The detailed view does show a detection of the 3.3~\um~PAH feature in the WFSS spectrum, while the MRS spectrum is too noisy for a clear detection - a demonstration of the fact that the WFSS mode has improved continuum sensitivity than the MRS mode in the wavelengths of overlap. The continuum flux is well matched between MRS and WFSS in both cases, indicating that the wavelength calibration inaccuracy has a modest impact on the overall spectrophotometric calibration in this portion of the array, as suggested in Section~\ref{subsec:wfss_extraction}.

\section{Final sample}\label{sec:sample}

\begin{figure}
    \centering
    \includegraphics[width=\linewidth]{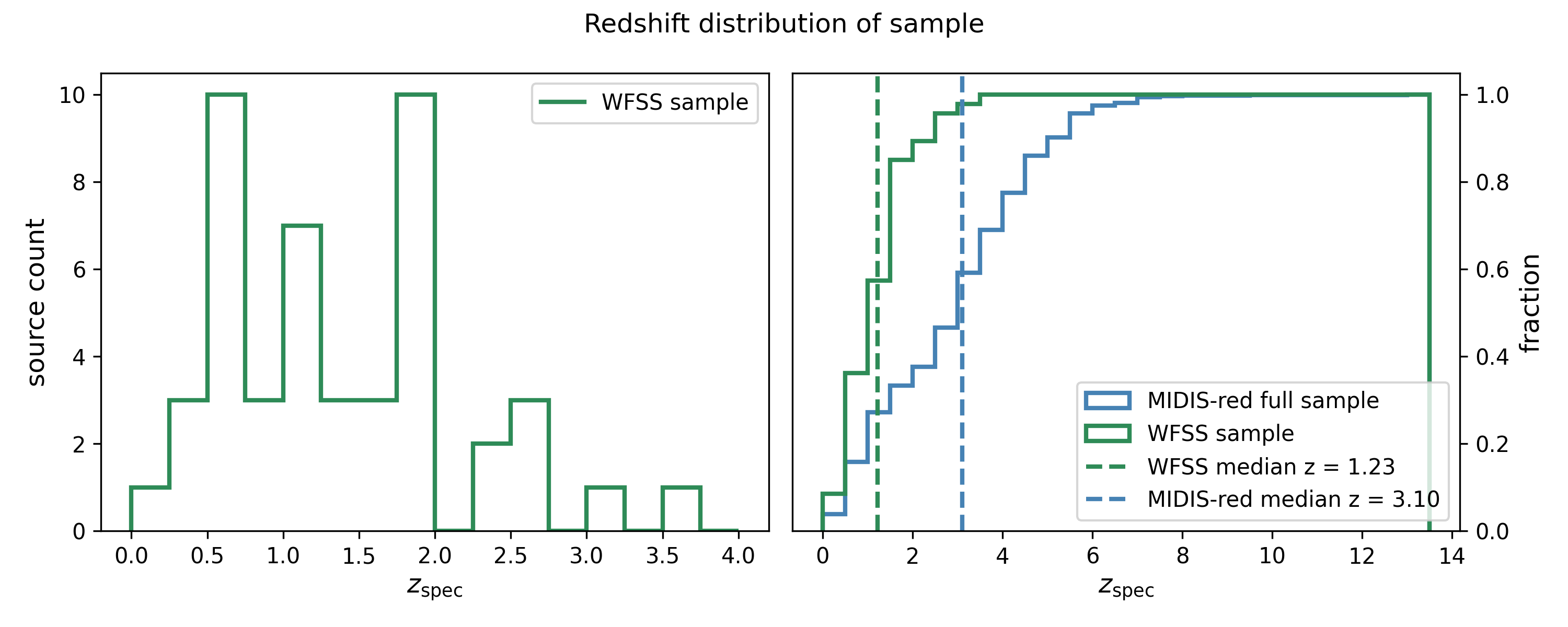}
    \caption{Distribution of spectroscopic redshifts in our sample. See Table~\ref{tab:wfss_sources} for redshifts of individual galaxies and relevant literature references. (L) simple histogram of the number counts vs. redshift in our WFSS sample. (R) comparison of the WFSS sample with the full MIDIS-red catalog of 2600+ sources with confirmed spectroscopic redshifts.}
    \label{fig:specz_dist}
\end{figure}

The final sample was further curated to eliminate sources which are too close to the detector edge, those which do not have spectroscopic redshifts or counterparts in the MIDIS or JADES photometric catalogs. In Table~\ref{tab:wfss_sources} we list the full set of 47 sources with simple identifiers unique to our program (based on their x, y locations in the field), their MIDIS/JADES counterparts, MIDIS coordinates and spectroscopic redshifts. The median spectroscopic redshift of the sample is 1.23; the distribution of redshifts is shown in Fig.~\ref{fig:specz_dist}. The figure also shows a comparison of the $z_{\mathrm{spec}}$ distribution of our sample with that of the full MIDIS-red imaging source catalog (Melinder et al, in prep), which contains 2622 sources with confirmed $z_{\mathrm{spec}}$ in the HUDF. Note that the full MIDIS survey has a larger spatial coverage than the area covered by our observations, and goes substantially deeper; this is not a ``like for like'' comparison. RGB near-infrared imaging from the JADES survey~\citep{2026ApJS..283....6E} is shown alongside the spectra for each source in Fig.~\ref{fig:all_sources_1}. We note that a larger sample can likely be extracted from this dataset given improved calibration and de-blending methods; for this work with the tools available, we use conservative selection methods. Our dataset is ``unbiased'' insofar as being SNR-limited in our F560W direct images. 

The HUDF is an extremely well-studied field, and many galaxies in our sample have ancillary data across the spectrum. In Table~\ref{tab:wfss_sources} we indicate the sources that were covered in the ALMA ASPECS survey and follow-up programs~\citep{Walter2016, Dunlop2017, boogaard2019, Boogaard2024}, as well as those included in~\citet{2025A&A...704A.100G}'s sample of X-ray sources. Of the 16 galaxies with X-ray counterparts, 12 are identified as AGN in~\citet{2025A&A...704A.100G}. Two sources (src 23 and 28 in Table~\ref{tab:wfss_sources}) are members of the HUDF46 Hubble-identified protocluster at $z \sim$ 1.84~\citep{2015ApJ...804..117M}, recently studied further with JWST by~\citet{2026arXiv260224162R}. One source is identified with $z_{\mathrm{spec}}$ of 0.001, i.e. this is a local stellar interloper. The highest redshift galaxy in the sample (source 39 in Table~\ref{tab:wfss_sources} and Fig~\ref{fig:all_sources_1}) at $z=$ 3.712 is an ALMA-detected X-ray AGN with associated Ly-alpha emission in MUSE~\citep{Aravena2020, Boogaard2024}.

\section{3.3~\um~PAH emitters in the HUDF}\label{sec:pah_emitters}

\subsection{Source identification and spectral fits}\label{subsec:pahfits}

For this work we focus on identifying and analyzing the galaxies in our sample that show PAH emission, in particular the 3.3~\um~feature which falls within the WFSS bandpass for 0.67 $< z <$ 3.1. This assumes a long-wavelength cutoff of 13.5~\um; theoretically the WFSS coverage extends to 14~\um. Of the sample of 47, 31 galaxies fall into the redshift range for detection of the 3.3~\um~PAH feature. Seven of these are known AGN (see Table~\ref{tab:wfss_sources}). We select the PAH emitters using a 2-step fitting process. First we perform a continuum-only and a continuum + Drude profile fit to the spectrum, isolating the spectral region around the expected PAH feature location. In both cases we fit the continuum with a first-order polynomial. Visual inspection of the fit and comparing the continuum + Drude against the continuum-only fit and residuals, we determine if the PAH feature is detected. The galaxies with a detected PAH feature are then passed to a more detailed fitting routine.

Of the 31 galaxies in the 0.67 $< z < $3.1 redshift range, a 3.3~\um~PAH feature was detected in 19 of the spectra, ranging in $z$ from 0.954 to 2.642. Four of these have known X-ray counterparts, three of which have been identified as AGN~\citep{2017ApJS..228....2L, 2025A&A...704A.100G} (sources 14, 28 and 47). For the other sources in the redshift range the spectrum was either too low SNR or the feature was not detected. The continuum was fit using a linear fit model in the adjacent line-free region, as in the first fitting stage. Given the uncertainty in wavelength calibration accuracy, we perform the fit in a relatively wide window (estimated line centre $\pm$ 2~\um). We allow the line centroid to vary by $\pm$1~\um~from the predicted observed wavelength, and the FWHM by $\pm$30\% from an initial estimate of 0.4~\um. These detailed fits include a Monte Carlo error analysis, drawing random samples from the uncertainties on the spectra (n=200) to obtain the mean best-fit parameters and their uncertainties. This fitting method gives a reasonable estimate of uncertainty on the fit, however it does not account for several systematic sources. 

The first is the calibration uncertainty, described in Section~\ref{subsec:wfss_extraction}. The median difference between the best-fit line centroid and the expected observed wavelength is 3 nm, but the spread over all sources is wide, ranging from -720 nm to +400 nm. For 15 of 19 sources with 3.3~\um~PAH detections, the wavelength offset is $<$ 50 nm. It is however difficult to relate these offsets directly to calibration accuracies, as shifts in centroids may also originate in the spatial distribution of the emission in the galaxies themselves (many of which are spatially resolved). Nonetheless, the galaxies with the largest offsets are src 17/JADES205597 ($z_{\mathrm{spec}} = 1.883$, $\Delta \lambda \simeq -720$ nm) is the galaxy closest to the right detector edge in our PAH emitter sample; and src 46/JADES207057 ($z_{\mathrm{spec}} = 1.922$, $\Delta \lambda \simeq +400$ nm) is located in the Lyot part of the imager field, near the upper edge of the detector and one of the furthest removed from our calibration reference location at the slit centre. 

\begin{figure}
    \centering
    \includegraphics[width=\textwidth]{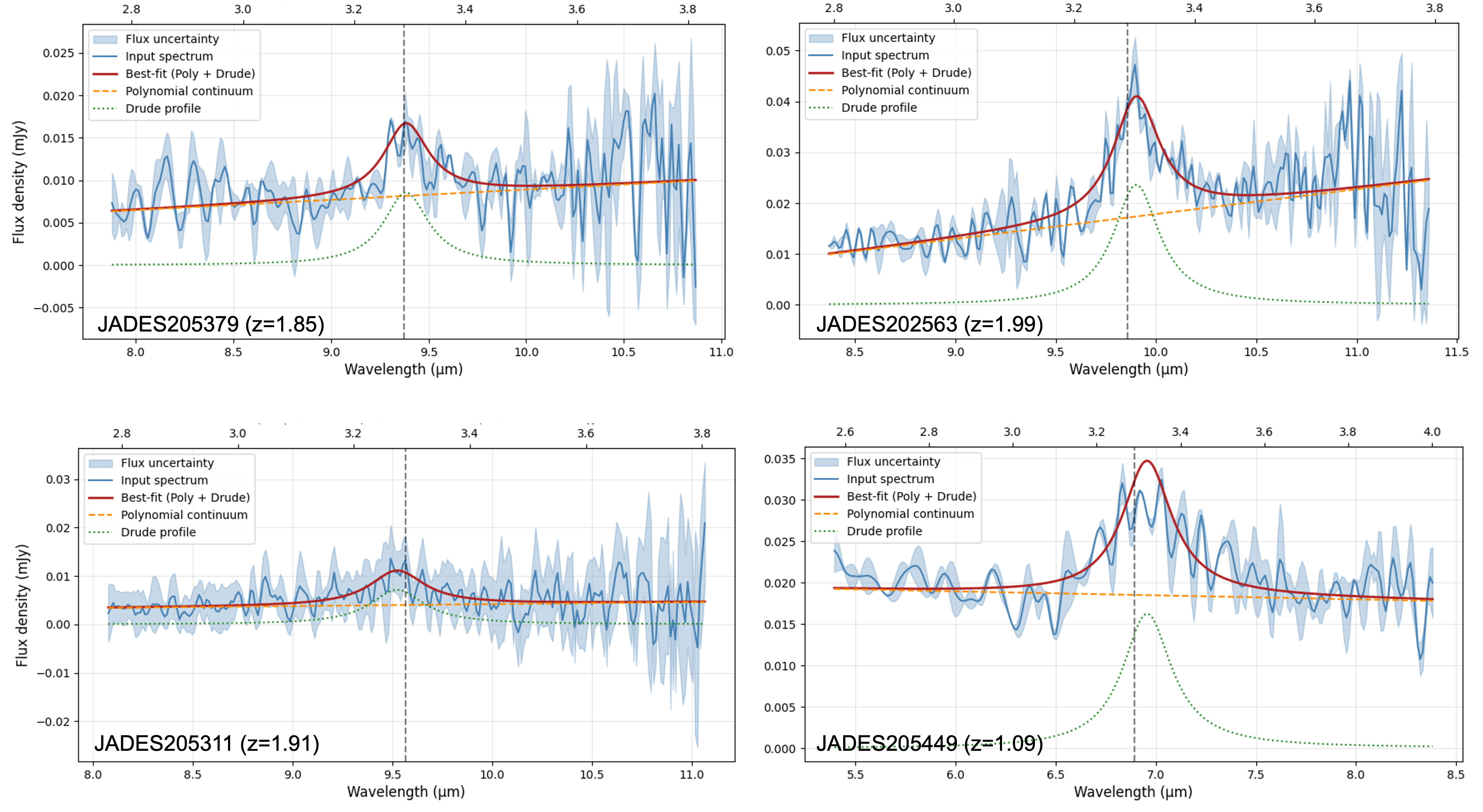}
    \caption{Examples of fits to the 3.3~\um~PAH feature in our sample of galaxies. Each plot shows the best 1st-order polynomial fit to the continuum, the best-fit Drude profile, and the combined best fit to the spectral region around the feature. Uncertainties on the spectrum flux are shown in the blue shaded region. x-axes show both the observed and rest wavelengths (bottom and top axes, respectively). The top-right panel shows the fit for source ALMA-C10, discussed in detail in Section~\ref{subsec:alma10}; this spectrum also shows a clear detection of the 3.4~\um~aliphatic feature. }
    \label{fig:pahfits_examples}
\end{figure}

The details of the fitting method bring additional sources of uncertainty; \citet[their Figure 10]{mckinney2026} investigate several aspects. First, the 3.3~\um~PAH feature has a weaker satellite feature, the 3.4~\um~aliphatic line, as well as a broad H$_2$O ice absorption feature around 3~\um. They find the inclusion of the aliphatic feature in the fit, modeled as a second Drude profile, has a negligible impact on the measurement of $L_{\mathrm{3.3}}$; however, fitting the water ice absorption feature is reported to affect the measured $L_{\mathrm{3.3}}$ by $\sim$0.1 dex. \citet[their Figure 6]{2026PASJ...78..454K}, in a sample of $z<$ 0.4 galaxies from \textit{AKARI}, model the 3.4~\um~aliphatic feature more comprehensively with a 4-component model (3.41 - 3.56~\um), finding a more significant aliphatic contribution to the total PAH flux, albeit with a large scatter ($\sim$5-90 \%).  While these features are visible in the highest-SNR spectra in our sample (see e.g. source 208/JADES202563/ALMA-C10 in the top-right panel of Fig.~\ref{fig:pahfits_examples} foran example of a clear 3.4~\um~feature detection), we did not routinely include them in our fits for this initial demonstration of WFSS capabilities. We add a 0.1 dex uncertainty to the errors on $\log L_{3.3}$ to account for the presence of the ice feature. \citet{rigopoulou_polycyclic_2024, 2025A&A...696A.135G} also provide a detailed discussion of the role of ice absorption and extinction in the estimation of PAH fluxes and ratios.  We also do not perform any extinction correction for our PAH fluxes. From the best-fit PAH profiles we compute the 3.3~\um~ PAH luminosities using:

\begin{equation}
L_{3.3} = P_{3.3} \times 4\pi \times d^2
\end{equation}

where $P_{3.3}$ is the 3.3~\um~PAH flux and $d$ is the luminosity distance. The measured line fluxes and luminosities are listed Table~\ref{tab:pahfits}, and example plots of fits are shown in Fig.~\ref{fig:pahfits_examples}. 


\subsection{SED fitting}\label{subsec:sed_fits}


We used {\sc bagpipes} \citep{2018MNRAS.480.4379C} to perform SED fitting and derive the stellar properties of our PAH emitters. We note that all PAH emitters in our sample have photometric counterparts in the MIDIS and SMILES catalogs \citep{midis2025,smiles2024}, providing HST~\citep{2019ApJS..244...16W}, JWST-NIRCam, and JWST-MIRI photometry and thus a comprehensive wavelength coverage from $0.4$ to $25.5\,\mu$m. We did not incorporate far-IR flux densities. Given that some of the sources are spatially extended, we adopted Kron photometry~\citep{1980ApJS...43..305K}, which more effectively captures the total emission our sources~\citep[e.g.][]{2026arXiv260115956R}. In all cases, the redshift was fixed to the spectroscopic value. Note that given the calibration issues with this particular dataset, we did not attempt full spectrophotometric SED fitting, using only the photometric points. 

{\sc bagpipes} relies on synthetic stellar population templates from \citet{bruzual_stellar_2003} with a \citet{2001MNRAS.322..231K} IMF, adopting an upper-mass cutoff of $100\,M_{\odot}$, while nebular emission is modeled with {\sc cloudy} \citep{2013RMxAA..49..137F}. We employed a continuity non-parametric SFH model \citep{2019ApJ...876....3L}. The age-bin edges, defined in look-back time from the redshift of observation, were determined individually for each source based on its redshift and distributed logarithmically between $z=30$ and the age of the Universe at the source redshift. We adopted the same approach for the age parameter.

Stellar masses were allowed to vary between $10^{5}$ and $10^{13}\,M_{\odot}$, adopting a uniform prior in logarithmic space. Dust attenuation was modeled using the Calzetti reddening law \citep{2000ApJ...533..682C}, with $A_V$ allowed to vary between 0 and 6. The stellar metallicity was also left free, spanning from sub-solar to super-solar values, while the ionization parameter was fixed to $\log_{10}(\mathcal{U})=-2$. We did not include an AGN component in the SED modeling, as our primary aim is to characterize the stellar populations and global stellar properties of the sources.

Finally, we compute the total infrared luminosity, $L_{\mathrm{IR}}$, by integrating the best-fit SED model from 8 to 1000~\um, following~\citet{2014MNRAS.445.1598H}.

\section{Results}

\begin{figure}
    \centering
    \includegraphics[width=\textwidth]{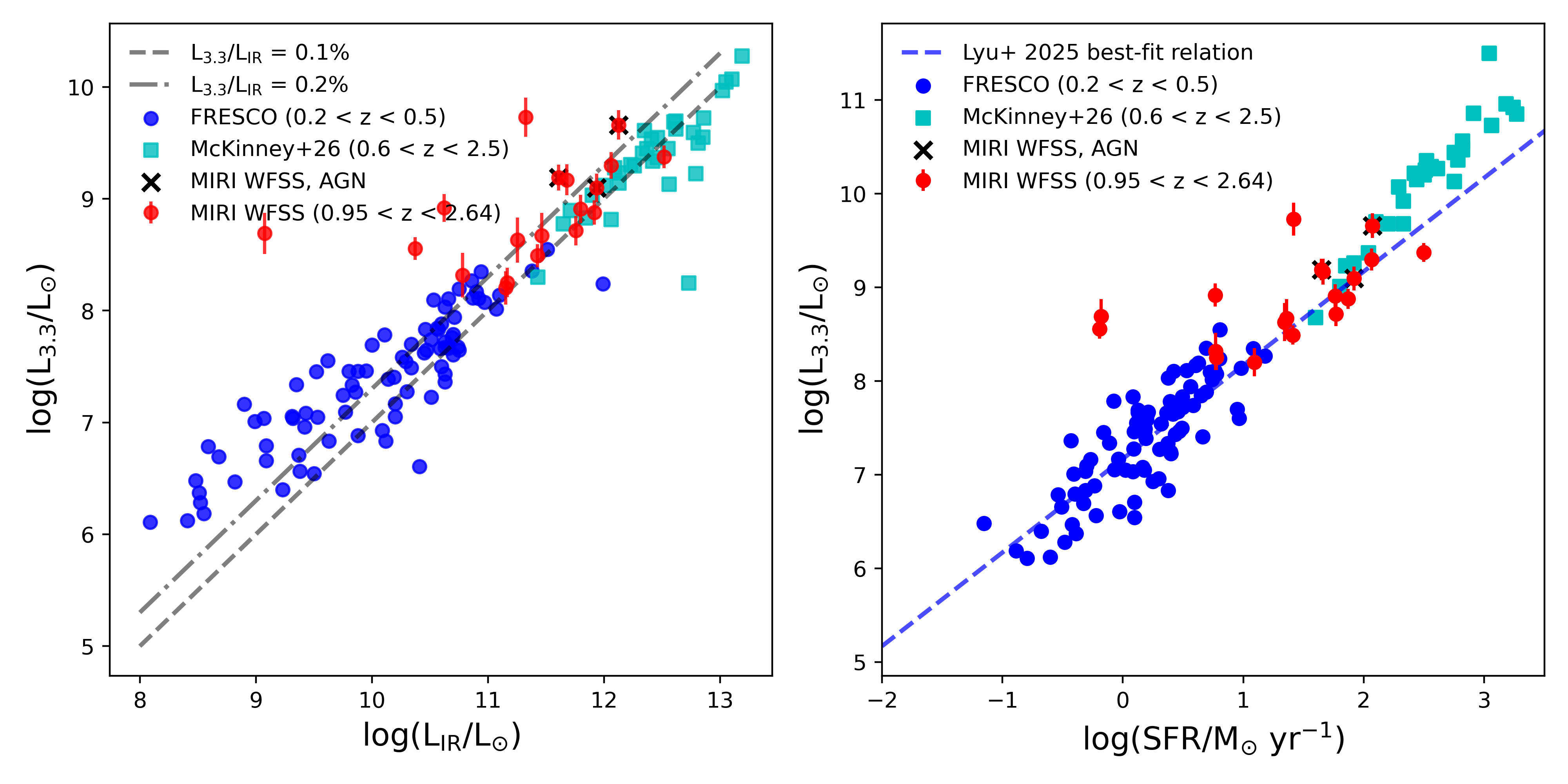}
    \caption{The 3.3~\um~PAH luminosity $L_{3.3}$ plotted against the galaxies' total IR luminosity ($L_{\mathrm{IR}}$) (L) and the star formation rate as derived from the best-fit SED models (R). The new MIRI WFSS measurements are shown in red, and compared against recent literature results, notably from~\citet{lyu2025} using data from FRESCO survey at lower $z$, and from~\citet{mckinney2026} based on MIRI LRS slit spectroscopy. The three AGN in our sample have black x's overlaid on the red symbols. The best-fit relations derived in these publications are also shown.}
    \label{fig:loglogplots}
\end{figure}

The 3.3~\um~PAH feature fits combined with the SED fit results allow us to plot the correlation of $L_{\mathrm{3.3}}$ against the galaxies'  total infrared luminosity $L_{\mathrm{IR}}$ and the SED-derived star formation rate. We plot these correlations in Fig.~\ref{fig:loglogplots}. The plots compare our measurements against two literature references described in the Introduction:  \citet{lyu2025} use NIRCam/grism spectra from the FRESCO survey~\citep[0.2$<z<$0.5]{fresco2023}; and~\citet{mckinney2026} who obtained MIRI LRS fixed-slit spectra of 37 Spitzer-selected Luminous Infrared Galaxies (LIRGS), i.e. biased towards very luminous systems (log $L_{\mathrm{IR}}/L_{\odot} > 11.5$) in the range 0.6 $<z<$ 2.5 . Note that the Figure shows~\citet{mckinney2026}'s data points fitted with a very similar methodology as our own - the dataset termed ``clip'' in their paper, rather than the ``decomp'' points which are the results of a more comprehensive fitting method (including e.g. extinction correction). 

Our HUDF sample extends~\citet{lyu2025}'s data to higher $L_{\mathrm{IR}}$ and higher redshifts; compared with~\citet{mckinney2026}'s sample the HUDF galaxies overlap significantly in redshift and luminosity but extend to lower-luminosity objects. 
 
Our data shows good agreement with the L$_{3.3}$/L$_{\mathrm{IR}}$ ratio of 0.1-0.2\% reported by~\citet{2020ApJ...905...55L} and~\citet{lyu2025}, trending to the higher value. The median L$_{3.3}$/L$_{\mathrm{IR}}$ ratio in our sample is 0.17\%, with 10th and 90th percentiles of 0.09\% and 2.5\%, respectively. We note that the scatter is clearly biased towards higher values.\citet{lyu2025} tentatively find a change in the $\log L_{3.3}$/$L_\mathrm{IR}$ ratio with $L_\mathrm{IR}$, shifting from $\sim$0.1\% at $L_\mathrm{IR} \sim 10^{11}-10^{12.5} L_{\odot}$ to  $\sim$0.2\% at $L_\mathrm{IR} \sim 10^{9}-10^{10} L_{\odot}$. Our sample largely consists of galaxies in the $L_\mathrm{IR} \sim 10^{11}-10^{12.5} L_{\odot}$ range, so our median ratio of 0.17\% is high compared to their range. The three AGN in the PAH emitter sample are marked with an additional black x in Fig.~\ref{fig:loglogplots}; their $\log L_{3.3}$/$L_\mathrm{IR}$ ratios are in line with the full sample. We note however that the 3.3~\um~PAH detection rate for the AGN in the sample, 3/7 or $\sim$40\%, is lower than the rate for the full sample (19/31 or $\sim$60\%). This may point to a systematically lower PAH flux in AGN compared with star-forming galaxies. Data calibration quality would have to be improved and a larger sample gathered to assess whether these findings are due to systematics or measurement issues, or physically meaningful.


The right-hand panel of Fig.~\ref{fig:loglogplots} plots the $L_{\mathrm{3.3}}$ measurements against the SED-fit-derived SFRs. The calibration of this relation from the local Universe to Cosmic Noon is of great interest given that the PAH feature provides a more direct measurement of dust-obscured star formation, which is dominant in this era. Our sample follows the best-fit relation from~\citet[0.2$< z <$0.5]{lyu2025}, again extending their data to higher redshifts and higher star formation rates, and providing an unbiased method of studying dust-obscured star formation at these redshifts. The scatter is however substantial; this is likely due to calibration issues affecting the PAH flux measurements. The outliers in the $L_{\mathrm{3.3}}$ vs. L$_{\mathrm{IR}}$ relation are likely affected by calibration issues. For the most significant outlier, on examination suffers from poor background subtraction due to a bright neighbour, and is located near the edge of the field, likely causing larger-than-average calibration errors. The best-fit SED model also shows a poor fit to the mid-IR photometry. 

\begin{table}
\centering
\begin{tabular}{lcccccccc}
\hline\hline
Source ID & MIDIS ID & $z_\mathrm{spec}$ & $P_{3.3}$ & $\log L_{3.3}/L_\odot$ & $L_\mathrm{IR}$ & $L_{3.3}$/$L_\mathrm{IR}$ & $\log \mathrm{SFR}_\mathrm{SED}$ \\
 & & & ($10^{-20}$\,W\,m$^{-2}$) & & ($10^{11}\,L_\odot$) & & ($M_\odot\,\mathrm{yr}^{-1}$) \\
\hline
3 & MIDIS033236.40-274747.48 & $1.7669$ & $7.35 \pm 1.71$ & $8.63 \pm 0.20$ & $1.79^{+0.14}_{-0.17}$ & 0.0024 & $1.34$ \\
12 & MIDIS033236.20-274726.43 & $1.5531$ & $8.51 \pm 0.05$ & $8.56 \pm 0.10$ & $0.24^{+0.04}_{-0.04}$ & 0.0152 & $-0.19$ \\
14$^{A}$ & MIDIS033235.83-274718.91 & $1.9080$ & $22.10 \pm 0.80$ & $9.19 \pm 0.12$ & $4.05^{+0.23}_{-0.21}$ & 0.0038 & $1.65$ \\
15 & MIDIS033236.97-274727.20 & $1.9100$ & $28.20 \pm 1.14$ & $9.30 \pm 0.12$ & $11.53^{+0.68}_{-0.76}$ & 0.0017 & $2.06$ \\
17 & MIDIS033235.52-274715.95 & $1.8828$ & $12.20 \pm 0.66$ & $8.92 \pm 0.12$ & $0.42^{+0.08}_{-0.08}$ & 0.0198 & $0.77$ \\
19 & MIDIS033238.56-274730.62 & $2.6421$ & $34.40 \pm 5.85$ & $9.73 \pm 0.17$ & $2.11^{+0.36}_{-0.43}$ & 0.0253 & $1.42$ \\
23 & MIDIS033238.80-274714.93 & $1.8497$ & $11.60 \pm 0.21$ & $8.88 \pm 0.11$ & $8.25^{+0.47}_{-0.65}$ & 0.0009 & $1.87$ \\
25 & MIDIS033239.89-274715.27 & $1.0953$ & $45.70 \pm 2.70$ & $8.91 \pm 0.13$ & $6.29^{+0.42}_{-0.36}$ & 0.0013 & $1.76$ \\
26 & MIDIS033238.49-274702.65 & $0.9536$ & $14.20 \pm 1.08$ & $8.25 \pm 0.13$ & $1.46^{+0.09}_{-0.07}$ & 0.0012 & $0.78$ \\
28$^{A}$ & MIDIS033237.35-274645.70 & $1.8462$ & $19.30 \pm 1.23$ & $9.10 \pm 0.13$ & $8.61^{+0.61}_{-0.71}$ & 0.0014 & $1.92$ \\
30 & MIDIS033239.82-274653.77 & $1.0965$ & $8.97 \pm 1.04$ & $8.20 \pm 0.15$ & $1.42^{+0.09}_{-0.11}$ & 0.0011 & $1.09$ \\
36 & MIDIS033241.84-274657.20 & $1.9982$ & $18.80 \pm 1.64$ & $9.17 \pm 0.14$ & $4.79^{+0.28}_{-0.36}$ & 0.0031 & $1.66$ \\
37 & MIDIS033241.69-274655.67 & $1.9960$ & $5.95 \pm 1.44$ & $8.67 \pm 0.21$ & $2.91^{+0.26}_{-0.20}$ & 0.0016 & $1.36$ \\
43 & MIDIS033243.63-274658.93 & $1.5693$ & $12.00 \pm 0.86$ & $8.72 \pm 0.13$ & $5.71^{+0.39}_{-0.35}$ & 0.0009 & $1.77$ \\
46 & MIDIS033244.21-274700.94 & $1.9224$ & $2.90 \pm 0.66$ & $8.32 \pm 0.20$ & $0.60^{+0.06}_{-0.06}$ & 0.0034 & $0.77$ \\
47$^{A}$ & MIDIS033241.02-274631.62 & $2.4540$ & $35.10 \pm 2.49$ & $9.66 \pm 0.13$ & $13.32^{+0.49}_{-0.51}$ & 0.0034 & $2.07$ \\
202 & MIDIS033236.12-274744.76 & $1.4154$ & $14.60 \pm 2.83$ & $8.69 \pm 0.18$ & $0.01^{+0.01}_{-0.01}$ & 0.4112 & $-0.18$ \\
206 & MIDIS033237.60-274743.94 & $1.0966$ & $17.40 \pm 0.10$ & $8.49 \pm 0.10$ & $2.67^{+0.13}_{-0.13}$ & 0.0012& $1.41$ \\
208 & MIDIS033240.07-274755.65 & $1.9978$ & $30.10 \pm 0.08$ & $9.37 \pm 0.10$ & $32.96^{+1.61}_{-1.55}$ & 0.0007 & $2.50$ \\
\hline
\end{tabular}
\caption{List of sources with 3.3~\um~PAH detections, showing PAH fluxes ($P_{3.3}$), PAH luminosities ($L_{3.3}$), total infrared luminosities ($L_{\mathrm{IR}}$), $L_{3.3}/L_{\mathrm{IR}}$ ratio, and star formation rates (SFR). Source ID's tagged $A$ are the 3 AGN in the sample. $L_{\mathrm{IR}}$ and SFR are derived from the best-fit SED models.}
\label{tab:pahfits}

\end{table}

\section{Discussion}\label{sec:discussion}

Our observations have yielded the first mid-infrared dispersed map of the HUDF. In a modest amount of time (6 hours, including overheads), we obtained spectra of 47 sources with known counterparts in catalogs from the MIDIS, SMILES and JADES surveys, 14 additionally with known ALMA counterparts, 12 AGN; all with known spectroscopic redshifts. For many of these galaxies, this is the first time ever mid-infrared spectra have been obtained, providing a wealth of new information. 

\subsection{Data quality}\label{subsec:data_quality}

The data presented here were obtained when no calibration pipeline was available for MIRI WFSS data. Sources were identified in a direct image, positions translated to the dispersed images, where cutouts were produced for each source. Spectra were extracted using simple boxcar extraction, and calibrations applied from publicly available calibration reference files for the MIRI LRS mode, which uses the same detector and the same disperser. By examining the current calibration reference files, we estimated the potential wavelength calibration offsets and their impact on the flux calibration of the spectra (see Figs.~\ref{fig:dispersion_comp} and \ref{fig:dphotom_test}). These figures are limited to comparison of the two well-calibrated positions in the field, and we note that the location of the slitless LRS mode is not very accessible for WFSS observations, given its location near the coronagraphic subarrays. The comparison is therefore not fully representative of our data. Nonetheless, examining the distortion differences between these locations gives a reasonable estimate of the expected uncertainties in our dataset. Estimating the wavelength calibration uncertainties in our dataset, even for a sample with high-confidence spectroscopic redshifts, is challenging for three reasons: (i) many galaxies in our sample are spatially resolved, with the spatial light profile along the dispersion axis convolved into the line profile; (ii) the PAH features themselves are intrinsically broad features, spectrally resolved, with an adjacent satellite feature at 3.4~\um, which is itself likely a blend of several components; and (iii) a prominent H$_2$O ice feature which is also known to affect the PAH fit if not accounted for. Despite these complications, our fit results are consistent with the prediction that the distortion increases with distance from the calibrated location (in our case, the slit centre); we see the largest wavelength offsets ($\geq$0.5~\um) in sources near the right-most edge of the detector array. 

The background subtraction method using fixed-aperture source masking proved effective and required no additional background observations or calibration reference files. Some regions of over- or under-subtraction do however remain, especially in regions where targets are closely spaced, and this was noted to affect the continuum levels in the final extracted spectra. This can be prevented with improved masking of close-neighbouring spectra during the spectral extraction, and careful cleaning or masking of regions of poor subtraction. Scaling a master background may give better results given that absolute background levels can vary by a few percentage points between dither positions; crowded regions will likely also benefit from such an approach, as creating a ``clean'' background from a crowded field may be challenging.  

In addition, we did not attempt to de-blend overlapping spectra, instead masking the affected regions of the spectra. The 2-observation strategy with 5 degree PA angle offset did minimize the number of sources that were fully blended with nearby neighbours; in most cases at least partial spectra could be recovered from the observation at the offset PA. Contamination mitigation in WFSS spectra is commonly done using a forward modeling approach, using known spectral trace information; in due course such tools will be adapted to the MIRI WFSS mode, and/or integrated into the JWST calibration pipeline. Even in the absence of de-blending methods, performing observations at two (or more) position angles allows the retrieval of additional sources - this is the recommended practice. A related complication is the spectral contamination of sources found in close proximity, even when not overlapping. Bright and/or extended sources may contaminate fainter neighbours; this could be exacerbated by the short-wavelength detector scattering effects known to produce the `cruciform artifact'' in MIRI imaging point spread functions~\citep{2021PASP..133a4504G}.

With the WFSS mode now formally in routine operation (since July 2026), the JWST calibration pipeline will provide improved calibration methods and, in due course, decontamination tools. Sophisticated community-developed tools are also available for optical/NIR slitless spectroscopy reduction and analysis, which can be adapted to work with MIRI WFSS data~\citep[e.g.][]{brammer_2023_8370018, 2023ApJ...953...53S}. With improved tools and calibration, we estimate an additional $\sim$10 sources could be retrieved from our dataset.

\subsection{MIRI WFSS for the study of galaxies at Cosmic Noon and beyond}\label{subsec:cosmicnoon}

Defining the Cosmic Noon era broadly as the $\sim$3.5 Gyr period between $z = \sim$ 1--3, we find around 60\% of our WFSS sample lies within this redshift range. Only 2 sources are at $z >$ 3, the remainder are at lower redshifts. The 3.3~\um~PAH feature, which we focus on in this work, is covered by the WFSS wavelength range from 0.67 $< z < $ 3.1, assuming a long-wavelength cutoff at 13.5~\um. The 6.2 and 7.7~\um~PAH features are accessible up to redshifts of 1.18 and 0.75, respectively, assuming the same cutoff. For high-SNR sources , there is a redshift range around $z \sim$ 0.7--0.75 where all 3 PAH features can be detected in a single spectrum. In addition, the WFSS covers the Pa-$\alpha$ line from $z$ of 1.9 to 5.4; the Br-$\alpha$ line from $z$ of 0.4 to 2. Fern\'{a}ndez Aranda et al (2026, in prep) demonstrate a wider range of investigations on a similar but deeper dataset, covering a broader redshift range.

The MIRI MRS mode provides wavelength coverage to $\sim$28.5~\um~and therefore is able to detect (at least theoretically) the 3.3~\um~ and other PAH features over a much wider range of redshifts, and at higher spectral resolving power of R$\sim$1500-3500. The PAHSPECS program recently demonstrated the capability of the MRS in this area~\citep{2026arXiv260618244D, 2026arXiv260618230L}. The extended wavelength coverage and spatial mapping capability however comes at the expense of observational efficiency: the MRS requires three grating settings, i.e. three exposures, to cover the full wavelength range. The WFSS mode has better (continuum) sensitivity than than the MRS in its shorter wavelength range\footnote{https://jwst-docs.stsci.edu/jwst-mid-infrared-instrument/miri-performance/miri-sensitivity}; and its field of view is larger. While the MRS mode provides rich spatially-resolved spectroscopy, slitless spectroscopy modes can also provide spatial information, albeit at lower resolution. \citet{nelson2024} detected the signature of rotation in the H$\alpha$ line profile of a z $> 5$ galaxy in slitless spectra from JWST-NIRCam. Using a combination of slitless spectra, imaging and modeling, \citet{2025ApJ...991..188E} created high-fidelity H$\alpha$, [SII] and [SIII] emission line maps for a sample of 20 galaxies at 0.6 $< z <$ 1.35 to study their spatially resolved star formation properties and metallicities. The larger field of view of the WFSS mode, its sensitivity and 5-14~\um~coverage in a single exposure makes WFSS an efficient mode for targeting larger, statistically significant samples in unbiased (or less-biased) surveys; particularly in fields with existing high-quality photometric coverage. The data quality demonstrated in this work for Cosmic Noon-era galaxies shows how the MIRI WFSS mode is ideally suited for surveying galaxies in this redshift regime; potentially extending to higher redshifts for sufficiently bright targets and supported by mature calibration and analysis pipelines.  

The 3.3~\um~PAH feature is of considerable interest, as a star formation rate indicator particularly sensitive to dust-obscured star formation, which dominates the star formation budget in this era. Our sample of HUDF galaxies at Cosmic Noon shows $L_{\mathrm{3.3}}$ following the $L_{\mathrm{3.3}}$ vs. $L_{\mathrm{IR}}$ relation established at lower redshifts in the NIR (trending towards the higher end of the 0.1 - 0.2\% ratio), and in matched redshift range from a targeted survey with MIRI LRS~\citep{lyu2025, mckinney2026}; as well as the low-redshift $L_{\mathrm{3.3}}$ vs. SFR relation~\citep{lyu2025}. Tightening this relation with improved calibration accuracy and larger samples will further strengthen the predictive power of $L_{\mathrm{3.3}}$ for galaxies' SFR across cosmic time. The ability to perform surveys with reduced bias using a WFSS mode also allows for testing of theories of the dependence of the 3.3~\um~PAH emission on galaxies' physical conditions and environment, testing e.g. correlations with metallicity observed locally with JWST (and earlier missions) to higher redshifts~\citep{rigopoulou_polycyclic_2024, 2026arXiv260805286K, 2026ApJ...997...20G}. 

In addition, the 3.3~\um~PAH feature is an important diagnostic for the properties of dust and its evolution over cosmic time. Originating from the smallest dust grains, the strength of the feature, in particular in relation to that of the 6.2 and 7.7~\um~features where these are simultaneously accessible, can constrain ionization properties of dust and grain size distribution~\citep{2020MNRAS.494..642M, 2023MNRAS.524.3429M}. In this work we study only the 3.3~\um~PAH feature, but the WFSS wavelength coverage provides redshift ranges where multiple PAH bands are detectable, allowing a more detailed study of band ratios. The presence and strength of the aliphatic satellite feature at 3.4~\um~provide a measure of the aliphatic fraction of the PAHs.~\citet{lyu2025} find a correlation of this fraction with SFR indicators, showing a decrease with increasing SFR or $L_{\mathrm{IR}}$ suggesting UV radiation may preferentially destroy aliphatic bonds in PAH molecules. Some detections of the 3.4~\um~satellite feature are evident in our sample, however given the high calibration uncertainties and the weakness of the feature we do not attempt to test this correlation in our sample. With the WFSS mode's ability to efficiently cover large fields, it provides an opportunity to gather a larger sample that bridges the gap between low redshifts and Cosmic Noon, in a variety of environments.

\section{Conclusion}

In this work we have shown the first mid-infrared slitless spectra taken with the $R \sim$ 100 MIRI WFSS observing mode in the Hubble Ultra Deep Field. With a total on-sky exposure time of 3.9 hours, we extract and calibrate spectra of 47 unique sources, all known from previous photometric and spectroscopic surveys, with $z_{\mathrm{spec}}$ ranging 0.32 to 3.7 (excepting the single foreground star, src 10) from a median $z_{\mathrm{spec}}$ of 1.23. We find the following results:

\begin{enumerate}
\item{Using custom data reduction code and calibration references files developed for the existing LRS (slit and slitless) modes, we extract and calibrate the spectra of our sources across all exposures in the program. The spectra are combined, masking regions of poor quality or contamination from neighbours. Based on 3.3~\um~PAH feature fits on a subset of 19 galaxies, we find the majority ($\sim$80\%) of recovered central wavelengths within 50 nm of their expected locations, with a median of 3 nm. Sources near detector edges show wavelength offsets of up to $\sim$0.7\um. A $\Delta \lambda$ of 50 nm implies a spectrophotometric calibration error around 10\%, the higher wavelength offsets can increase the error to 50\% or more, varying with wavelength. }
\item{Of 31 galaxies in our sample in 0.67 $< z < $3.1, where the 3.3~\um~PAH feature falls in the WFSS wavelength range, we detect the PAH feature in 19 sources, or 62\%. Of the seven known AGN in this sample of 31, just three (43\%) have 3.3~\um~PAH detections. Our sample shows strong complementarity with datasets from \citet{lyu2025} and \citet{mckinney2026}, extending the former's redshift range beyond $z \sim$ 0.5, and the latter's sample to lower luminosities at similar redshifts.}.
\item{Using best-fit models to the PAH feature + continuum and SED fits to the photometry from ancillary surveys, we plot the PAH luminosity $L_{\mathrm{3.3}}$ against the total infrared luminosity and SED-derived star formation rate. We find a median $L_{\mathrm{3.3}}/L_{\mathrm{IR}}$ ration of 0.0017 (0.17\%), in line with findings from the lower-$z$ sample of ~\citet{lyu2025}, who find a ratio in the range 0.1--0.2\%, and $z \sim$ 1--2 sample of ~\citet{mckinney2026} - albeit with considerable scatter. The $L_{\mathrm{3.3}}$ vs. SFR correlation shows similar agreement with published relations.}

\end{enumerate}

We conclude that MIRI WFSS is a powerful, efficient new mode for extragalactic surveys in the mid-infrared with JWST. Its wavelength coverage and sensitivity are ideally suited to observations of Cosmic Noon-era galaxies, covering a number of key spectral features in this redshift regime. At lower redshifts, WFSS can provide some spatially resolved information for extended targets; at higher redshifts, the mode will allow detection and characterization of the most luminous galaxies. Improvements in calibration and advanced analysis tools are under active development to fully leverage the capabilities of MIRI in this new mode.

\begin{table*}
\centering
\begin{longtable}{clllrrclc}
\hline\hline
WFSS ID & MIDIS ID & JADES ID & RA (deg) & Dec (deg) & $z_\mathrm{spec}$ & $z_\mathrm{spec}$ ref & ALMA ID & X$-$ray \\
\hline
3 & MIDIS033236.40$-$274747.48 & JADES203254 & 53.151674 & $-$27.796522 & 1.7670 & 1 & \nodata & \nodata \\
4 & MIDIS033235.38$-$274737.79 & JADES203937 & 53.147436 & $-$27.793831 & 1.2260 & 1 & \nodata & \nodata \\
5 & MIDIS033235.80$-$274735.01 & JADES204050 & 53.149168 & $-$27.793059 & 1.2230 & 1 & \nodata & 667$^{A}$ \\
6 & MIDIS033239.23$-$274758.61 & JADES202206 & 53.163439 & $-$27.799614 & 0.6650 & 1 & \nodata & \nodata \\
8 & MIDIS033237.63$-$274744.55 & JADES400532 & 53.156810 & $-$27.795709 & 1.0970 & 1 & \nodata & 709 \\
10 & MIDIS033238.02$-$274741.96 & JADES285600 & 53.158423 & $-$27.794988 & 0.0010 & 1 & \nodata & \nodata \\
11 & MIDIS033235.99$-$274725.96 & JADES204579 & 53.149961 & $-$27.790543 & 2.6320 & 2 & 1mm.C21 & \nodata \\
12 & MIDIS033236.20$-$274726.43 & JADES204578 & 53.150852 & $-$27.790675 & 1.5530 & 1 & \nodata & \nodata \\
13 & MIDIS033237.31$-$274729.64 & JADES204226 & 53.155471 & $-$27.791567 & 0.6690 & 1 & \nodata & 700$^{A}$ \\
14 & MIDIS033235.83$-$274718.91 & JADES205311 & 53.149276 & $-$27.788585 & 1.9080 & 1 & \nodata & 668$^{A}$ \\
15 & MIDIS033236.97$-$274727.20 & JADES333720 & 53.154034 & $-$27.790890 & 1.9100 & 3 & 1mm.C02 & \nodata \\
16 & MIDIS033236.89$-$274726.43 & JADES333721 & 53.153716 & $-$27.790676 & 1.3170 & 1 & \nodata & \nodata \\
17 & MIDIS033235.52$-$274715.95 & JADES205597 & 53.147997 & $-$27.787763 & 1.8830 & 1 & \nodata & \nodata \\
18 & MIDIS033238.78$-$274732.41 & JADES204131 & 53.161589 & $-$27.792337 & 0.4580 & 1 & 1mm.C30 & 724 \\
19 & MIDIS033238.56$-$274730.62 & JADES204449 & 53.160658 & $-$27.791838 & 2.6420 & 1 & \nodata & \nodata \\
20 & MIDIS033238.50$-$274725.57 & JADES204767 & 53.160432 & $-$27.790437 & 1.6120 & 1 & \nodata & \nodata \\
21 & MIDIS033240.68$-$274731.25 & JADES204264 & 53.169493 & $-$27.792013 & 0.3240 & 1 & \nodata & 754$^{A}$ \\
22 & MIDIS033237.74$-$274707.17 & JADES206205 & 53.157262 & $-$27.785324 & 0.6680 & 1 & 1mm.C32 & \nodata \\
23 & MIDIS033238.80$-$274714.93 & JADES205379 & 53.161683 & $-$27.787480 & 1.8500 & 1 & 1mm.C17 & \nodata \\
24 & MIDIS033240.33$-$274723.07 & JADES204984 & 53.168037 & $-$27.789741 & 0.6190 & 1 & \nodata & S-36 \\
25 & MIDIS033239.89$-$274715.27 & JADES205449 & 53.166193 & $-$27.787576 & 1.0950 & 1 & 1mm.C16 & 749 \\
26 & MIDIS033238.49$-$274702.65 & JADES206703 & 53.160389 & $-$27.784071 & 0.9540 & 1 & 1mm.C33 & \nodata \\
27 & MIDIS033239.64$-$274709.39 & JADES205964 & 53.165180 & $-$27.785942 & 1.3160 & 1 & \nodata & 745$^{A}$ \\
28 & MIDIS033237.35$-$274645.70 & JADES208134 & 53.155646 & $-$27.779361 & 1.8460 & 1 & 1mm.C18 & 703$^{A}$ \\
29 & MIDIS033238.80$-$274649.15 & JADES207671 & 53.161650 & $-$27.780318 & 0.6210 & 1 & \nodata & 727$^{A}$ \\
30 & MIDIS033239.82$-$274653.77 & JADES207501 & 53.165918 & $-$27.781603 & 1.0970 & 1 & 3mm.11 & \nodata \\
32 & MIDIS033240.97$-$274655.37 & JADES207375 & 53.170716 & $-$27.782047 & 0.7670 & 1 & \nodata & \nodata \\
33 & MIDIS033238.55$-$274634.60 & JADES209117 & 53.160608 & $-$27.776278 & 2.5540 & 1 & 1mm.C01 & 718$^{A}$ \\
34 & MIDIS033238.44$-$274632.16 & JADES209119 & 53.160180 & $-$27.775599 & 0.6230 & 1 & \nodata & \nodata \\
35 & MIDIS033238.25$-$274630.38 & JADES209126 & 53.159388 & $-$27.775106 & 1.2160 & 1 & \nodata & \nodata \\
36 & MIDIS033241.84$-$274657.20 & JADES207221 & 53.174324 & $-$27.782557 & 1.9980 & 1 & 1mm.C14b & \nodata \\
37 & MIDIS033241.69$-$274655.67 & JADES207227 & 53.173695 & $-$27.782132 & 1.9960 & 4 & 1mm.C14a & \nodata \\
38 & MIDIS033238.60$-$274631.53 & JADES209122 & 53.160835 & $-$27.775426 & 0.6220 & 1 & \nodata & \nodata \\
39 & MIDIS033238.03$-$274626.56 & JADES209777 & 53.158471 & $-$27.774045 & 3.7120 & 2 & 1mm.C08 & 715$^{A}$ \\
40 & MIDIS033241.43$-$274651.70 & JADES207590 & 53.172611 & $-$27.781029 & 0.6200 & 1 & \nodata & \nodata \\
41 & MIDIS033242.84$-$274702.77 & JADES206907 & 53.178497 & $-$27.784104 & 3.1890 & 1 & \nodata & 788$^{A}$ \\
42 & MIDIS033238.97$-$274630.46 & JADES209124 & 53.162367 & $-$27.775127 & 0.4190 & 1 & \nodata & \nodata \\
43 & MIDIS033243.63$-$274658.93 & JADES207079 & 53.181775 & $-$27.783037 & 1.5690 & 1 & \nodata & \nodata \\
44 & MIDIS033239.34$-$274623.86 & JADES210075 & 53.163918 & $-$27.773294 & 2.4490 & 1 & \nodata & 801$^{A}$ \\
45 & MIDIS033238.10$-$274614.10 & JADES210989 & 53.158770 & $-$27.770583 & 0.9970 & 1 & \nodata & \nodata \\
46 & MIDIS033244.21$-$274700.94 & JADES207057 & 53.184208 & $-$27.783594 & 1.9220 & 1 & \nodata & \nodata \\
47 & MIDIS033241.02$-$274631.62 & JADES209357 & 53.170910 & $-$27.775451 & 2.4540 & 2 & 1mm.C04 & 809$^{A}$ \\
202 & MIDIS033236.12$-$274744.76 & JADES203473 & 53.150495 & $-$27.795767 & 1.4150 & 1 & \nodata & \nodata \\
206 & MIDIS033237.60$-$274743.94 & JADES400532 & 53.156673 & $-$27.795538 & 1.0970 & 1 & \nodata & \nodata \\
208 & MIDIS033240.07$-$274755.65 & JADES202563 & 53.166956 & $-$27.798792 & 1.9980 & 1 & 1mm.C10 & \nodata \\
215 & MIDIS033237.41$-$274725.93 & JADES333719 & 53.155859 & $-$27.790535 & 0.6710 & 1 & \nodata & \nodata \\
224 & MIDIS033241.41$-$274717.46 & JADES205125 & 53.172549 & $-$27.788183 & 0.6220 & 1 & \nodata & \nodata \\
\hline
\end{longtable}
\caption{List of all WFSS sources in the sample. The WFSS ID is an arbitrary identification, based on the pixel location in the direct image field, from bottom-left to top-right; WFSS IDs $>$ 200 are sources identified only in Observation 2. Missing indices indicate sources removed from the sample as discussed in the text. $z_{\mathrm{spec}}$ sources: (1) MIDIS ~\citep{midis2025}, (2) FRESCO~\citep{fresco2023}, (3) NGDEEP~\citep{ngdeep2024}; (4) ALMA~\citep{boogaard2019, boogaard2020}. MIDIS $z_{\mathrm{spec}}$ are collected from the MUSE Ultra Deep Field~\citep{2023A&A...670A...4B}, 3DHST~\citep{2012ApJS..200...13B, 2016ApJS..225...27M} and JADES~\citep{2026arXiv260115956R}. ALMA IDs from the ASPECS survey~\citep{Walter2016}. X-ray source information and IDs from~\citet{2025A&A...704A.100G} ($^A$: AGN).}\label{tab:wfss_sources}
\end{table*}

\begin{figure}[p]
\centering
\includegraphics[height=0.95\textheight]{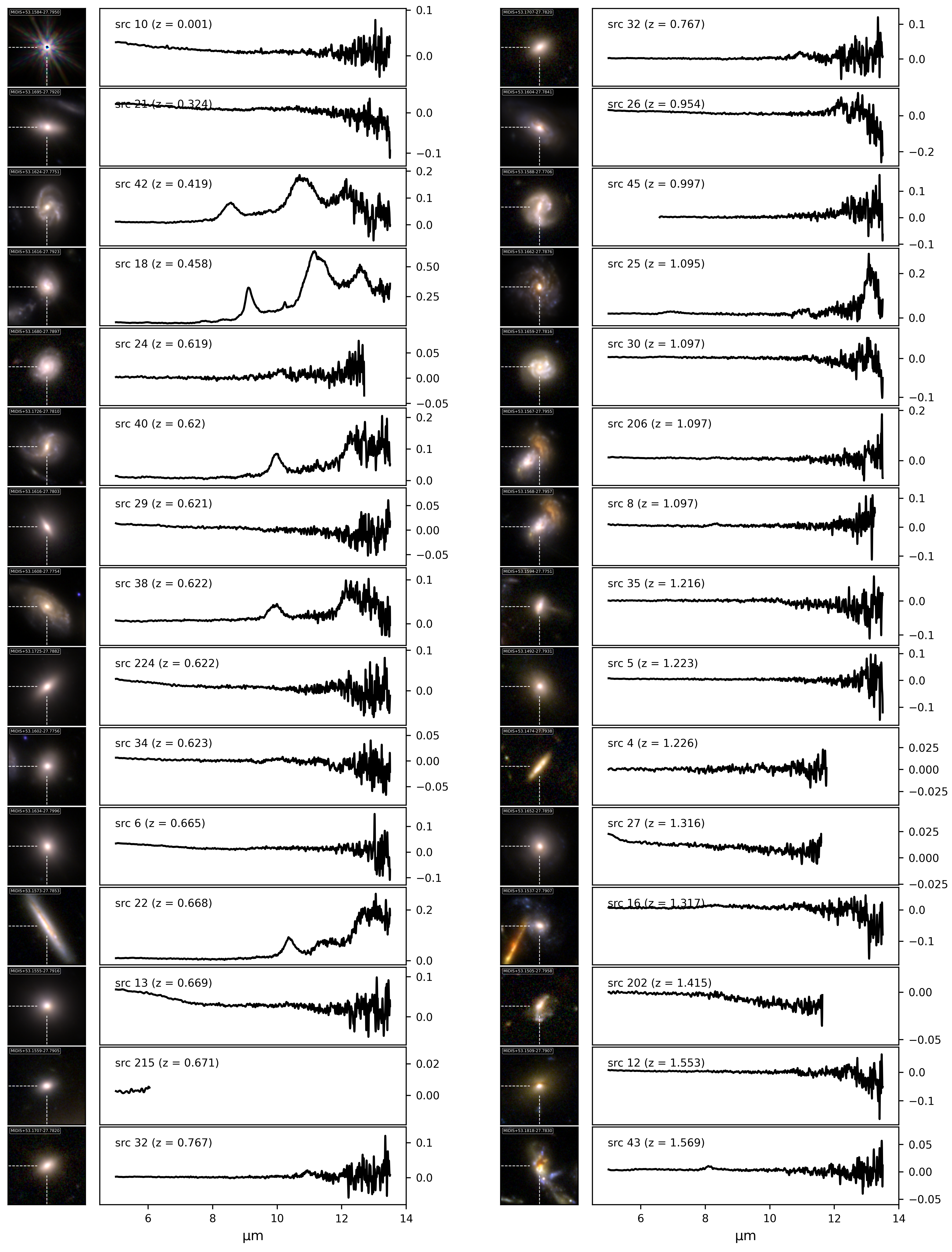}
\caption{RGB images and MIRI WFSS spectra for all sources, in order of ascending $z_{\mathrm{spec}}$. The RGB images measure 3 $\times$ 3'' and are constructed from NIRCam filters F090W, F150W and F200W. The dashed lines point to the centre coordinate of the WFSS source. The spectra are plotted in observed wavelengths, and fluxes in mJy.}\label{fig:all_sources_1}
\end{figure}

\begin{figure}[p]
\ContinuedFloat
\centering
\includegraphics[height=0.95\textheight]{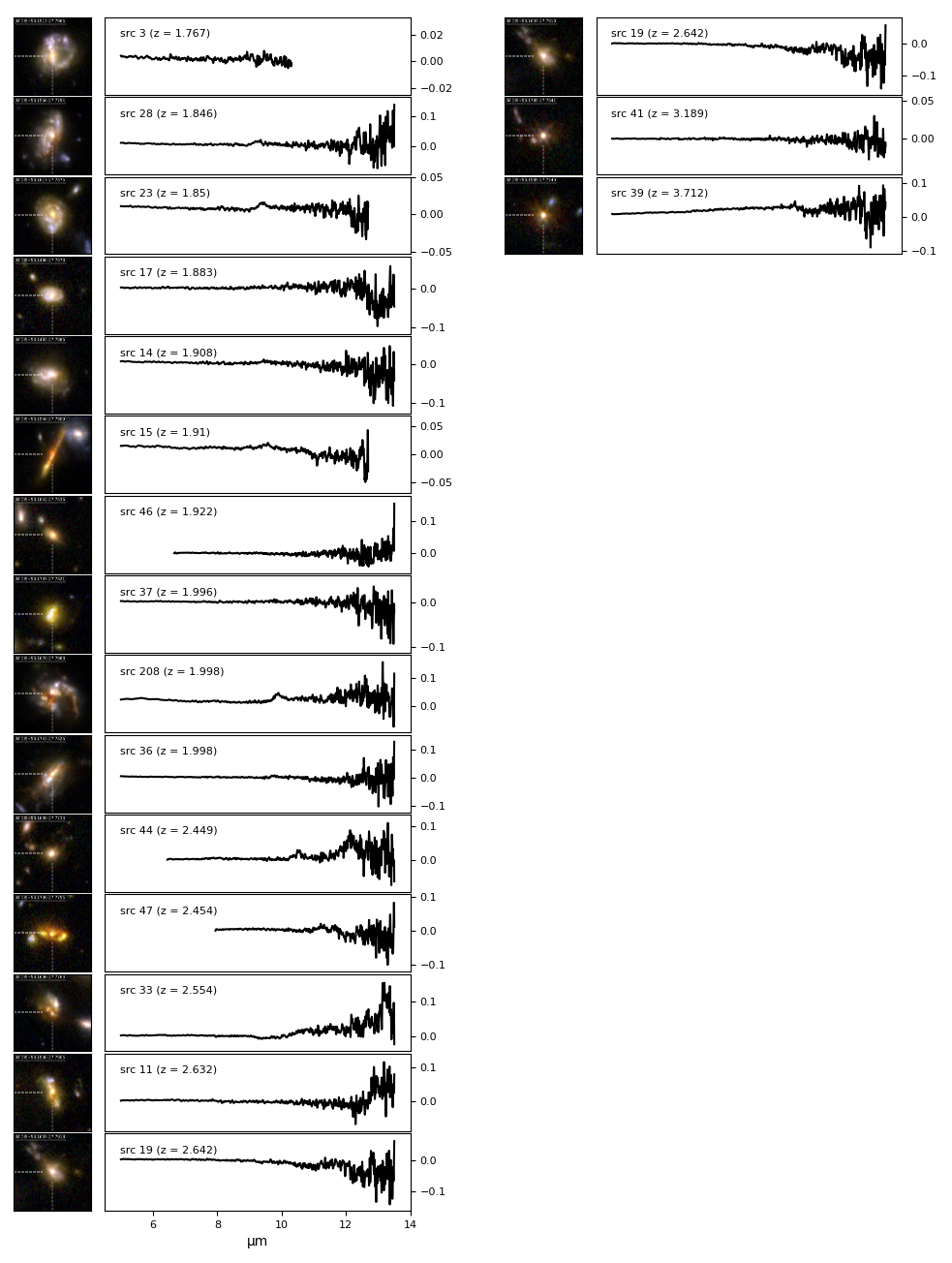}
\caption{Continued from previous page.}\label{fig:all_sources_2}
\end{figure}

\begin{acknowledgements}

The following software packages were used in this work: Astropy\footnote{https://www.astropy.org}, a community-developed core Python package and an ecosystem of tools and resources for astronomy \citep{astropy:2013, astropy:2018, astropy:2022}; Specutils~\citep{specutils2.2.0}; photutils~\citep{photutils2.3.0}; the JWST calibration pipeline~\citep{jwst1.17.1}; and other standard libraries. S.K. made use of a generative AI tool (Claude Sonnet 5) in the optimization of the code used to fit the spectral features, in some aspects of visualization (specifically, Fig.~\ref{fig:pahfits_examples}), and in the conversion of large tables into LaTeX format. No AI tools were used to develop original methods for calibration or analysis of the data, nor in the writing process of the manuscript or sourcing of references. All code can be shared on request, including any AI prompts used to generate code.

S.K. and R.F.A. acknowledge support from the European Space Agency through the Science Faculty - Funding reference ESA-SCI-E-LE-342. G\".{O}. and J.M. acknowledge support from the Swedish National Space Agency (SNSA) and the Swedish Research Council (VR). P.R. acknowledges support from the University of Texas at Austin Cosmic Frontier Center. L.A.B. acknowledges support from the Dutch Research Council (NWO) under grant VI.Veni.242.055 (\url{https://doi.org/10.61686/LAJVP77714}). A.A.H. acknowledges support from grant PID2021-124665NB-I00  funded by MCIN/AEI/10.13039/501100011033 and by ERDF A way of making Europe. K.I.C. acknowledges funding from the Dutch Research Council (NWO) through the award of the Vici Grant VI.C.212.036. S.G. and T.G. acknowledges financial support from the Cosmic Dawn Center (DAWN), funded by the Danish National
Research Foundation (DNRF) under grant DNRF140. J.P.P. acknowledges financial support from the UK Science and Technology Facilities Council, and the UK Space Agency, during part of this work. For the purpose of open access, the authors have applied a Creative Commons Attribution (CC BY) licence to the Author Accepted Manuscript version arising from this submission. D.L. is supported by the Carlsberg Foundation, grant CF25-0662. R.F.A. is supported by the European Research Council (ERC) under the European Union's Horizon 2020 research and innovation programme (DistantDust, Grant agreement No. 101117541).  

This work is based on observations made with the NASA/ESA/CSA James Webb Space Telescope.
 The work presented is the effort of the entire MIRI team and the enthusiasm within the MIRI partnership is a significant factor in its success.
 The following National and International Funding Agencies funded and supported the MIRI development: NASA; ESA; Belgian Science Policy Office (BELSPO); Centre Nationale d’Etudes Spatiales (CNES); Danish National Space Centre; Deutsches Zentrum fur Luftund Raumfahrt (DLR); Enterprise Ireland; Ministerio De Economia y Competividad; Netherlands Research School for Astronomy (NOVA); Netherlands Organisation for Scientific Research (NWO); Science and Technology Facilities Council; Swiss Space Office; Swedish National Space Agency (SNSA); and UK Space Agency.
 MIRI drew on the scientific and technical expertise of the following organizations: Ames Research Center, USA; Airbus Defence and Space, UK; CEAIrfu, Saclay, France; Centre Spatial de Li\`ege, Belgium; Consejo Superior de Investigaciones Cientficas, Spain; Carl Zeiss Optronics, Germany; Chalmers University of Technology, Sweden; Danish Space Research Institute, Denmark; Dublin Institute for Advanced Studies, Ireland; European Space Agency, Netherlands; ETCA, Belgium; ETH Zurich, Switzerland; Goddard Space Flight Center, USA; Institute d’Astrophysique Spatiale, France; Instituto Nacional de T\'ecnica Aeroespacial,Spain; Institute for Astronomy, Edinburgh, UK; Jet Propulsion Laboratory, USA; Laboratoire d’Astrophysique de Marseille (LAM), France; Leiden University, Netherlands; Lockheed Advanced Technology Center (USA); NOVA Opt-IR group at Dwingeloo, Netherlands; Northrop Grumman, USA; Max Planck Institut f \"ur Astronomie (MPIA), Heidelberg, Germany; Laboratoire d’Etudes Spatiales et d’Instrumentation en Astrophysique (LESIA), France; Paul Scherrer Institut, Switzerland; Raytheon Vision Systems, USA; RUAG Aerospace, Switzerland; Rutherford Appleton Laboratory (RAL Space), UK; Space Telescope Science Institute, USA; Stockholm University, Sweden; Toegepast- Natuurwetenschappelijk Onderzoek (TNOTPD), Netherlands; UK Astronomy Technology Centre, UK; University College London, UK; University of Amsterdam, Netherlands; University of Arizona, USA; University of Cardiff , UK; University of Cologne, Germany; University of Ghent; University of Groningen, Netherlands; University of Leicester, UK; University of Leuven, Belgium;  Utah State University, USA.

\end{acknowledgements}

\bibliography{miri_wfss}{}
\bibliographystyle{aasjournalv7}



\end{document}